\documentclass[12pt,preprint]{aastex}

\usepackage{graphicx}
\usepackage{amssymb}
\usepackage{threeparttable}

\newcommand{\oii}{[O {\sc ii}]}
\newcommand{\oiii}{[O {\sc iii}]}
\newcommand{\heii}{He {\sc ii}}
\mathchardef\mhyphen="2D

\slugcomment{Accepted for publication in MNRAS}

\shorttitle{Observations of RQ Quasar Feedback. I.}
\shortauthors{Liu et al.}

\begin{document}


\title{Observations of Feedback from Radio-Quiet Quasars: \\
I. Extents and Morphologies of Ionized Gas Nebulae}


\author{Guilin Liu\altaffilmark{1}}
\email{liu@pha.jhu.edu}
\author{Nadia L. Zakamska\altaffilmark{1}}
\author{Jenny E. Greene\altaffilmark{2}}
\author{Nicole P. H. Nesvadba\altaffilmark{3}}
\and
\author{Xin Liu\altaffilmark{4,5}}

\altaffiltext{1}{Department of Physics \& Astronomy, Johns Hopkins University, 3400 N. Charles St., Baltimore, MD 21218, USA}
\altaffiltext{2}{Department of Astrophysical Sciences, Princeton University, Princeton, NJ 08544, USA}
\altaffiltext{3}{Institut d'Astrophysique Spatiale, CNRS, Universit\'{e} Paris-Sud, 91405 Orsay, France} 
\altaffiltext{4}{Harvard-Smithsonian Center for Astrophysics, 60 Garden Street, Cambridge, MA 02138, USA}
\altaffiltext{5}{Einstein fellow}

\begin{abstract}

Black hole feedback -- the strong interaction between the energy output of supermassive black holes and their surrounding environments -- is routinely invoked to explain the absence of overly luminous galaxies, the black hole vs. bulge correlations and the similarity of black hole accretion and star formation histories. Yet direct probes of this process in action are scarce and limited to small samples of active nuclei. In this paper we present Gemini Integral Field Unit observations of the distribution of ionized gas around luminous, obscured, radio-quiet quasars at $z\sim0.5$. We detect extended ionized gas nebulae via \oiii$\lambda$5007\AA\ emission in every case, with a mean diameter of 28 kpc. These nebulae are nearly perfectly round, with H$\beta$ surface brightness declining $\propto R^{-3.5\pm 1.0}$. The regular morphologies of nebulae around radio-quiet quasars are in striking contrast with lumpy or elongated \oiii\ nebulae seen around radio galaxies at low and high redshifts. We present the uniformly measured size-luminosity relationship of \oiii\ nebulae around Seyfert 2 galaxies and type 2 quasars spanning six orders of magnitude in luminosity and confirm the flat slope of the correlation ($R_{\rm [O~{\scriptscriptstyle III}]}\propto L_{\rm [O~{\scriptscriptstyle III}]}^{0.25\pm 0.02}$). We propose a model of clumpy nebulae in which clouds that produce line emission transition from being ionization-bounded at small distances from the quasar to being matter-bounded in the outer parts of the nebula. The model -- which has a declining pressure profile -- qualitatively explains line ratio profiles and surface brightness profiles seen in our sample. It is striking that we
    see such smooth and round large-scale gas nebulosities in this
    sample, which are inconsistent with illuminated merger debris and
    which we suggest may be the signature of accretion energy
    from the nucleus reaching gas at large scales.

\end{abstract}




\section{Introduction}
\label{sec:intro}

Feedback from accreting black holes has become a key element in modeling galaxy 
evolution \citep{tabo93, silk98, spri05}. For one thing, regulatory mechanisms 
are needed to keep massive galaxies from forming too many stars and becoming 
overly massive or blue at late times \citep{thou95,crot06}. Black holes are 
natural candidates because even a small fraction of the binding energy of the
infalling material is in principle sufficient to liberate the galaxy-scale gas 
from the galaxy potential. The discovery that the masses of supermassive black 
holes in inactive galaxies strongly correlate with the velocity dispersions and 
masses of their hosts' stellar bulges \citep{mago98,gebh00,ferr00,trem02,marc03, 
hari04,gult09,mcco11} also suggests that the energy output of the black hole in its 
most active -- quasar -- phase must be somehow coupled to the gas from which 
the stars form \citep{hopk06}. The problem remains to find direct observational 
evidence of black hole-galaxy self-regulation and to obtain measurements of 
feedback energetics. 

In some situations, accretion energy clearly has had an impact on the 
large-scale environment of the accreting black hole. For instance, radio jet 
activity in massive elliptical galaxies and brightest cluster galaxies deposits 
energy into the hot gas envelope of the cluster \citep{mcna07, rand11}. Likewise, 
there is clear evidence that powerful radio jets entrain warm gas and carry significant 
amounts of material out of their host galaxies \citep{vanb86, tadh91, whit92, vill99, 
nesv06, nesv08, fu09}, and the observed entropy profiles of galaxy clusters 
can be explained by invoking a heating mechanism from their central radio-loud 
quasars to reduce the gas supply from the hot intra-cluster medium
\citep{scan04,voit05a,voit05b,Dona06}.
Nevertheless, as radio-loud quasars only represent a small
fraction ($\sim 10\%$) of the entire quasar population, invoking this mechanism 
as the primary mode of limiting galaxy mass would require all galaxies to have 
undergone a radio-loud phase -- a conjecture which lacks direct evidence and 
contradicts a theoretical paradigm in which radio-loudness is determined by 
the spin of the black hole \citep{tche10}.  

Nuclear activity is known to drive outflows on small scales \citep{cren03b, 
gang08}, perhaps via the radiation pressure of the quasar \citep{murr95,prog00}. Broad 
absorption-line troughs are seen in $\sim 10-20\%$ of luminous quasars 
\citep{reic03}, or alternatively the outflows are ubiquitous, and this frequency simply 
reflects their covering factors \citep{weym81, gall07, shen08}. The velocities in 
broad absorption lines can be up to $\sim 10,000$ km/s, suggesting that high energies 
are involved, but in practice velocity is the only physical parameter in such 
outflows which is directly measurable. Careful photo-ionization modeling of 
specific absorption features \citep{arav08} demonstrated that in several objects 
the outflow appears to extend out to several kpc from the nucleus and to carry 
a large amount of kinetic energy and momentum \citep{moe09, dunn10}. Such analysis 
requires high-quality ultra-violet spectra of special broad absorption-line quasars 
showing weak metastable transitions and is only possible for a handful 
of sources. 

Over the past few years, we have undertaken a spectroscopic campaign to use the 
spatial distribution and kinematics of ionized gas to search for any impact of 
the black hole on the galaxies on large scales \citep{gree11, gree12}. Our 
approach is to measure the spatial distribution and kinematics of ionized gas 
in quasars. While such work has been conducted by several groups since it was 
pioneered by \citet{stoc87} and \citet{boro85}, our approach is different in 
a couple of respects. First, we have pushed these observations to the highest 
luminosity quasars, because the efficiency of feedback inferred from simulations 
of galaxy formation rises steeply with black hole mass \citep[][although the most 
massive black holes are barely active, at least at low redshifts]{crot06}. Second, 
we focused our observations on obscured quasars in which circumnuclear material 
blocks the line of sight to the observer.

Such objects -- the luminous analogs of Seyfert 2 galaxies in the framework of 
the unification model \citep{anto93} -- remained unidentified in large numbers 
until the Sloan Digital Sky Survey (SDSS; \citealt{york00}). Now we have a sample 
at $z < 0.8$ selected based on emission line ratios and the \oiii$\lambda$5007\AA\ 
(hereafter \oiii) line luminosity \citep{zaka03} which comprises nearly 1000 objects 
\citep{reye08} with bolometric luminosities up to $10^{47}$ erg s$^{-1}$ \citep{liu09}. 
Extensive follow-up by our group and others using the Hubble Space Telescope (\emph{HST}; \citealt{zaka06}), 
\emph{Chandra} and \emph{XMM} \citep{ptak06,vign10,jia12}, \emph{Spitzer} \citep{zaka08}, 
spectropolarimetry \citep{zaka05}, \emph{Gemini} \citep{liu09}, \emph{VLT} 
\citep{vill11a,vill11b}, Calar Alto 3.5m \citep{hump10}, and the \emph{VLA} 
\citep{zaka04,lal10} yields a detailed description of the optically selected 
obscured quasar population and their host galaxies. Among other findings, at the 
highest luminosities, it has become clear that these objects constitute at least 
half of the quasar population \citep{reye08} and are therefore representative of 
``typical'' black hole activity. These objects allow us to observe extended 
emission line regions without the overwhelming glare of the quasar itself. There are 
even some theoretical suggestions that quasars experience outflows preferentially in 
the obscured phase \citep{hopk06}.

In this paper, we present Gemini-North Multi-Object Spectrograph (GMOS-N) Integral 
Field Unit (IFU) observations of a sample of fourteen quasars and the analysis of 
the spatial extents, morphologies, and physical conditions of their narrow emission 
line regions. In Section \ref{sec:data} we describe sample selection, observations, 
data reduction and calibrations. In Section \ref{sec:science}, we present maps of 
the ionized gas emission, in Section \ref{sec:discussion} we discuss physical 
conditions and morphologies of the nebulae and we summarize in Section \ref{sec:conclusions}.
We use a $h$=0.71, $\Omega_m$=0.27, $\Omega_{\Lambda}$=0.73 cosmology throughout 
this paper. Objects are identified as SDSS Jhhmmss.ss+ddmmss.s in Table 
\ref{tab1} and are shortened to SDSS Jhhmm+ddmm elsewhere. The rest-frame 
wavelengths of the emission lines are given in air. 

\section{Data and measurements}
\label{sec:data}

\subsection{Sample selection}
\label{sec:selection}

We study eleven radio-quiet type 2 quasars selected from the catalog by 
\citet{reye08} according to the following criteria:

\begin{enumerate}

\item We select the most luminous quasars in the catalog, with \oiii\ line 
luminosities $L_{\rm [O~{\scriptscriptstyle III}]}>10^{42.8}$ erg s$^{-1}$, 
corresponding to intrinsic luminosities $M_B<-26.2$ \citep{reye08}. 

\item Among these luminous sources, we select objects with the lowest redshifts 
($z=0.4$--0.6) in order to maximize the spatial information of our observations.

\item We require the radio flux at 1.4 GHz $L_{\rm 1.4GHz}<10$ mJy as determined 
by the FIRST survey \citep{beck95,whit97}. Applying the $K$-correction formula
in \citet{zaka04}, this corresponds to $\nu L_{\nu}<1.1\times10^{41}$ erg s$^{-1}$ 
at the 1.4 GHz rest-frame frequency for $z=0.5$ sources. This is a conservative 
threshold which puts our targets about one order of magnitude below the separation 
of the radio-loud and radio-quiet objects at these \oiii\ luminosities and at these 
redshifts \citep{xu99,zaka04}.

\end{enumerate}

We use the distribution of objects in the \oiii--$L_{\rm radio}$ plane for defining 
radio-loud, radio-intermediate and radio-quiet objects because unlike definitions 
based on continuum flux, the one based on \oiii\ is less sensitive to the optical 
type. \citet{xu99} adopt a separation line with 
$L_{\rm radio}\propto L_{\rm [O~{\scriptscriptstyle III}]}^{0.5}$, and following 
this definition we find that approximately 10\% of type 2 quasars qualify as radio-loud 
\citep{zaka04}, although our dynamic range is not sufficient to determine 
whether this fraction varies significantly with luminosity. Our adopted definition 
produces similar fractions of radio-loud sources among type 1 and type 2 quasars 
\citep{reye08}, but yields a lower radio-loud fraction than the traditional definitions 
based on a constant optical continuum to radio continuum ratio \citep[e.g.,][]{jian07}. 

In addition to the 11 radio-quiet objects, we observed three radio-loud type 2 quasars 
to provide a comparison sample for ionized gas 
morphologies as a function of jet activity. Two radio-loud targets, 
SDSS~J0807+4946 and SDSS~J1101+4004, were selected from the \citet{reye08} 
catalog according to criteria 1 and 2. The FIRST image centered on the position
of SDSS~J0807+4946 shows an FR II \citep{fana74} double-lobed radio galaxy oriented 
at 60$^{\circ}$ East of North. The maximal lobe distance from the host galaxy is 
$\sim$50\arcsec\ (330 kpc). While there is no apparent radio core component, the two 
lobes are symmetric around the SDSS position, and there is no other compelling candidate 
host galaxy for the radio source. Therefore, we identify SDSS~J0807$+$4946 with the 
radio source 87GB~0804$+$4955, which has a spectral index of 
$\alpha=-1.0$ ($F_{\nu}\propto \nu^{\alpha}$) between 80 cm and 6 cm \citep{beck91}.

The radio flux of SDSS J1101+4004 places it in the radio-intermediate regime
\citep{xu99,zaka04}, but the FIRST image clearly shows a pair of radio lobes
oriented at 112$^{\circ}$ East of North and symmetrically located at 
$\sim$40\arcsec\ (230 kpc) away from the nucleus, which is a point source
with a peak flux of 17.48 mJy/beam \citep{beck95}. Using the FIRST data,
we find the integrated flux to be 45 mJy and 21 mJy for the northwest and 
southeast radio lobes, respectively.

The third radio-loud target is 3C67 at $z$=0.311,
a compact steep-spectrum source with a total extent of $\sim3$\arcsec\ and 
with lobes oriented at 173$^{\circ}$ East of North in the 5.0 GHz radio band
\citep{erac04}. The optical spectrum of this object is consistent with
type 2 quasar classification \citep{spin06}, and its \oiii\ line luminosity
is slightly lower than the majority of the other objects (see Table~\ref{tab1}).
The \emph{HST}/STIS long-slit spectroscopy finds the \oiii\ line emission to extend 
out to 2.7 kpc away from the nucleus \citep{odea02}. Not enough information is available 
for a direct comparison between the sensitivity of these observations and ours. We 
find general agreement between the orientation of the extended emission, while our 
detection of the faint \oiii\ emission extends further out, presumably because of 
better sensitivity. 

\subsection{Observations and data reduction} 

We observed eleven radio-quiet obscured quasars and three radio galaxies 
(Table~\ref{tab1}) with GMOS-N IFU \citep{alli02} in December 2010 
(program ID: GN-2010B-C-10, PI: Zakamska). We use the two-slit mode that
covers a 5\arcsec$\times$7\arcsec\ field of view, translating to a physical 
scale of 30$\times$42 kpc$^2$ at $z=0.5$, the typical redshift of our objects. 
The science field of view is sampled by 1000 contiguous 0.2\arcsec-diameter 
hexagonal lenslets, and simultaneous sky observations are obtained by 500
lenslets located $\sim$1\arcmin\ away. 
In view of their respective redshifts, 13 objects were observed in 
the {\it i}-band (7060--8500 \AA) so as to cover the rest frame wavelengths 
$4100<\lambda<5200$ \AA\ and thus H$\beta$ and \oiii. Of those, two were also 
observed in the {\it r}-band ($5620<\lambda<6980$ \AA) to map out 
\oii$\lambda$3727\AA. The last source, 3C67, is at a lower redshift, so we 
observed it only in the {\it r}-band for the \oiii\ coverage. We used the 
R400-G5305 grating leading to a dispersion of 0.687 \AA\ per pixel during the 
{\it i}-band observations and 0.680 \AA\ for the {\it r}-band. 

For each object in each band, we took two dithered exposures of 1800 sec each 
with an 0\farcs5 offset along the direction of the longer (7\arcsec) side 
of the rectangular field of view, with the exception of SDSS~J1101+4004 
where only one 1800 sec exposure was obtained. 
The two exposures that go into each observation are reduced separately. We 
perform the data reduction using the Gemini package for IRAF\footnote{The 
Image Reduction and Analysis Facility (IRAF) is distributed by the National 
Optical Astronomy Observatories which is operated by the Association of 
Universities for Research in Astronomy, Inc. under cooperative agreement with 
the National Science Foundation.}, following the standard procedure for GMOS 
IFU described in the tasks {\sl gmosinfoifu} and 
{\sl gmosexamples}\footnote{http://www.gemini.edu/sciops/data/IRAFdoc/gmosinfoifu.html},
except that (a) we use an overscan instead of a bias image throughout the
data reduction, and adjust the relevant parameters so that the bias correction
is applied only once on each image, and (b) we set the parameter ``weights''
of {\sl gfreduce} to ``none'' (in contrast to ``variance'' as suggested by
the standard example) to avoid significantly increased noise in some parts
of the extracted spectrum.
 
For each dither position, the routine {\sl gfreduce} trims, overscan-subtracts 
and extracts Gemini Facility Calibration Unit (GCAL) and twilight flat 
observations, and {\sl gfresponse} makes response curves with twilight 
correction. We similarly process the copper-argon arc calibration images 
using {\sl gfreduce} and then perform wavelength calibration using 
{\sl gswavelength} and {\sl gftransform}. 
We use {\sl gfreduce} to trim and overscan-subtract science frames, but 
then we interrupt the standard pipeline to remove the cosmic rays using the 
spectroscopy version of the {\sc L.A.Cosmic} software \citep{vdok01}. The 
cleaned output is then sent back to the master task {\sl gfreduce} for further 
processing: the data from separate CCDs are mosaiced and flat-fielded, the 
traces of the GCAL flat are referenced for extracting the spectra, the 
wavelengths of the spectra are calibrated, and the derived sky spectrum is 
subtracted. As the final step, the reduced data from each exposure are resampled 
and interpolated onto a data cube with a spatial sampling scale of 0\farcs1, 
and the two frames are shifted and combined to produce the final science 
data cube.

\subsection{Calibration} 
\label{sec:calib}

We flux-calibrate our data using the spectra of our science targets from 
the SDSS Data Release 7\footnote{http://www.sdss.org/dr7}. SDSS spectra 
are collected by fibers with a 3\arcsec~diameter at a typical seeing of 
$\sim$2\arcsec, and SDSS spectrophotometry is better than 10\% over its 
entire wavelength coverage of 3900--9100\AA\ \citep{abaz09}. 
Since SDSS fiber fluxes are calibrated using point spread function
(PSF) magnitudes, SDSS spectrophotometry is corrected for fiber losses.

We subtract the host galaxy continuum from both the SDSS and the IFU 
observations and concentrate on calibrating the \oiii\ observations. 
In order to do this, we first locate the pixel (the 0.1\arcsec\
``spaxel'') in which the \oiii\ line has the broadest wing on the 
spectrum. Then we manually select a pair of wavelength intervals on 
both sides of \oiii\ where the continuum is free of any line 
emission and artifacts (e.g., chip gaps, strong residuals of 
sky line removal). This pair of wavelength intervals is then fixed 
for all the pixels, and we linearly interpolate between them
to define the continuum, which we then subtract the best-fit 
line for each pixel. 

The seeing of our IFU observations varied between 0\farcs4 and 0\farcs7 
during the three observing nights. To simulate the SDSS fiber observations, 
we convolve the IFU image at each wavelength with a Gaussian kernel 
whose Full Width at Half Maximum (FWHM) satisfies 
$\rm FWHM^2+seeing^2={2\arcsec}^2$ to mimic the SDSS observing 
conditions and then we extract the spectrum using a 3\arcsec-diameter 
circular aperture. We then collapse the spectrum within the 
wavelength intervals used for local continuum 
fitting. The resultant [O {\sc iii}] intensity, when compared to 
the flux measured from the SDSS spectrum whose local continuum
under \oiii\ is subtracted in the same way as we do for the IFU data, provides
a flux calibration factor for all emission lines. The detailed velocity structure of the [O {\sc iii}] emission line is highly consistent between the 3\arcsec\ extraction from the IFU data (after spatial smoothing) and the SDSS fiber spectrum once the SDSS vacuum wavelengths are properly recalculated in the air. 

The calibration of the SDSS data includes a PSF correction to recover the flux outside the fibers assuming a 2\arcsec\ seeing, which needs to be removed for our purpose. In the last step of our calibration, we downgrade the image of our standard star to 2\arcsec\ resolution and calculate the correction factor for a 3\arcsec\ circular aperture centered on the star, which changes the final flux calibration by 7\%. We then take this into account for the final calibration of the IFU data against the SDSS spectra. 

\section{Measurements of the ionized gas quasar nebulae}
\label{sec:science}

\subsection{\oiii\ maps and surface brightness profiles} 
\label{sec:psf}

\begin{figure*}
\centering
\includegraphics[scale=0.85,clip,trim=0cm 0cm 0cm 0cm]{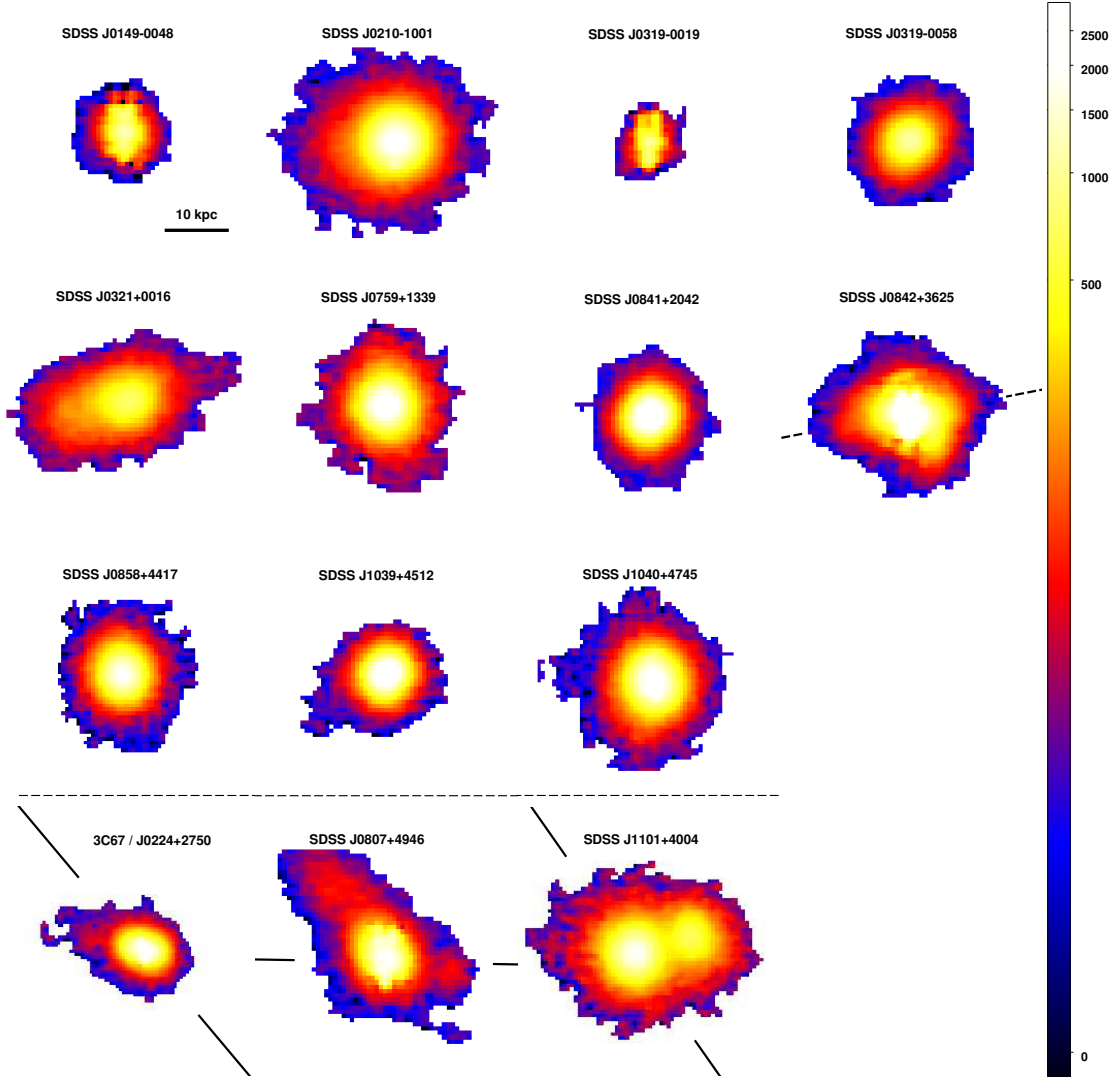}
\caption{Intensity maps of the \oiii\ line from the sample 
obscured quasars (11 radio-quiet and 3 radio-loud/radio-intermediate 
objects, separated by the dashed line), with a cutoff at a signal-to-noise 
ratio of 5, shown on a logarithmic color scale. The tick marks of the 
color bar are in units of $10^{-17}$ erg s$^{-1}$ cm$^{-2}$ arcsec$^{-2}$. 
All maps have the same spatial scale as shown in the top left panel. 
The orientation of the scattering axis for SDSS~J0842+3625 
\citep[101$^{\circ}$ east of north,][]{zaka05} is marked by thick dashed lines. 
The orientation of the jets from the radio objects is denoted by 
thick solid lines.} 
\label{fig:OIII}
\end{figure*}

We obtain a flux-calibrated \oiii\ line intensity map by collapsing 
the uncalibrated continuum-free IFU datacube over the \oiii\
wavelength range and multiplying it by the calibration factor 
described in Section~\ref{sec:calib}. The \oiii\ maps of the eleven 
radio-quiet type 2 quasars are shown in Figure \ref{fig:OIII}, where 
the false color is used to represent the intensity on a logarithmic 
scale. We zoom in and out on the objects depending on their redshifts 
to display all images on the same linear scale. We determine the noise 
level in each pixel on the map using wavelengths outside but close to 
the \oiii\ line. The surface brightness sensitivity (r.m.s. noise) of 
our \oiii\ maps is in the range 
$\sigma=1.0$--$2.2\times10^{-17}$ erg s$^{-1}$ cm$^{-2}$ arcsec$^{-2}$. 
We use a 5--$\sigma$ threshold to create these maps. 

\begin{figure*}
\centering
\includegraphics[scale=0.5,clip,trim=0cm 0cm 0cm 0cm]{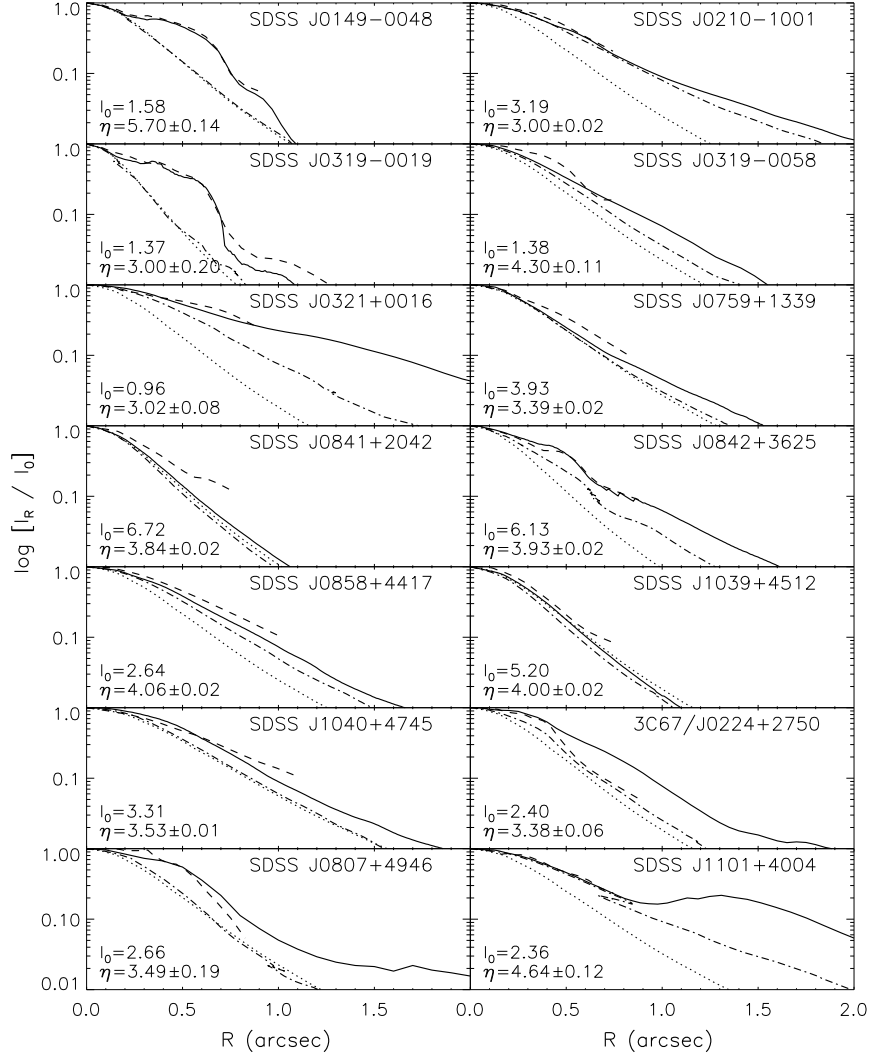}
\caption{\oiii\ surface brightness profiles of the quasars in our sample 
along both the semi-major axes of the best-fit isophotal ellipses (solid 
lines) and the semi-minor axes (dot-dash lines), compared to the profiles 
of continuum emission along the semi-major axes (dashed lines) and the 
PSF at the time of the observation (dotted lines). The peak \oiii\ surface 
brightness ($I_0$) marked on each panel is in $10^{-14}$ erg s$^{-1}$ 
cm$^{-2}$ arcsec$^{-2}$. The continuum emission profiles are truncated at 
their respective 5--$\sigma$ levels. Also given in each panel is the 
absolute value of the best-fit exponent ($\eta$) of power-law fits to 
the outer regions of the nebulae along the semi-major axes 
(i.e., $I_{R,\rm [O~{\scriptscriptstyle III}]}\propto R^{-\eta}$).} 
\label{fig:psf}
\end{figure*}

The average seeing of 0.4--0.7\arcsec\ corresponds to a typical linear 
resolution of $\sim$3 kpc at the typical redshifts of our sample. In Figure 
\ref{fig:psf} we show a detailed comparison of the \oiii\ surface brightness 
profiles with the PSFs. We fit elliptical isophotes to 
the continuum-free \oiii\ line intensity map of each object using the 
{\sl ellipse} task from the IRAF STSDAS package. In the Figure, we show 
the surface brightness of each isophote as a function of both semi-major 
and semi-minor axes. 
The seeing is determined by averaging the directly measured FWHM 
of a sample of field stars in the acquisition image taken right before 
the science exposure. In Figure~\ref{fig:psf}, each PSF is represented 
by the radial profile of our standard star whose width is scaled to 
match the seeing of each observation. The standard star is slightly
elongated (ellipticity = 0.07), and here we conservatively use the
profile along its major-axis. The radial profiles of our targets are
clearly extended except SDSS J0841+2042 and SDSS J1039+4512. These
two objects, although they appear only marginally resolved, demonstrate 
evident velocity gradient and \oiii\ line width variation across their 
maps (which we will show in an upcoming paper) and a strong radial change 
of \oiii/H$\beta$ line ratio similar to that seen in other objects (see Section~\ref{sec:line_ratio}). We thus 
conclude that all the quasar nebulae are spatially resolved by our IFU 
observations.

\subsection{Size Measurements}
\label{sec:size}

The physical size of a quasar's narrow-line region is one of the basic 
parameters that characterize the impact of the active nucleus on its 
environment; yet this value is not an easy one to define. The options
include (but are not limited to):

\begin{enumerate}

\item $R_{5\sigma}$, the radius of the best-fit ellipse which encloses
pixels with a signal-to-noise ratio of 5 or higher. This measurement
faithfully reports the observed extent of a quasar nebula, but the 
size depends on the depth that
an observation reaches, making it difficult to compare objects observed 
with different sensitivity and/or at different redshifts.

\item $R_{\rm eff}$, the effective radius which encloses half of the total 
luminosity of the line emission. Due to the steep decline of the light 
profiles in the central parts, this approach has the advantage that $R_{\rm eff}$ 
is relatively insensitive to the detection threshold or to faint extended 
emission. However, this is also the disadvantage if faint extended emission 
is what we are after. Furthermore, since there is no universal surface 
brightness profile amongst quasar nebulae, $R_{\rm eff}$ does not provide 
a complete characterization of the nebula.

\item $R_{\rm obs}$, the isophotal radius at an observed limiting surface 
brightness of $10^{-16}$ erg s$^{-1}$ cm$^{-2}$ arcsec$^{-2}$. In this
work, this threshold is chosen so that the measurements stay above $\sim3\sigma$ 
sensitivity level for our sample objects. This definition may be useful for 
nearby objects in which cosmological dimming effects are negligible. 

\item $R_{\rm int}$, the isophotal radius at an intrinsic limiting surface 
brightness of $10^{-15}$ erg s$^{-1}$ cm$^{-2}$ arcsec$^{-2}$, which is 
corrected by a factor of $(1+z)^4$ (ranging from 4.5 to 7.4 for our objects) 
in order to account for the cosmological surface brightness dimming effect. 
This threshold corresponds to a 6--$\sigma$ detection for the worst case in 
our sample. This relatively conservative threshold is a compromise between 
our data and the data with lower sensitivity and/or at higher redshifts 
which we take from the literature for various comparisons.
$R_{\rm int}$ is better physically motivated than the previous 
measures, in that it does not depend on the depth of the observations
and the redshifts of the objects.

\end{enumerate}

All the above radii are defined as the semi-major axes of the best-fit
ellipses at their respective isophotes or signal-to-noise ($S/N$) levels. The measured diameters of the ionized gas nebulae range from 15 to 40 kpc, as seen from the $R_{5\sigma}$ values determined from the \oiii\ maps. Not just the faint extended gas envelopes, but the bright central parts are resolved by our observations: the effective radii are from 1.4 to 2.9 times the effective radii of the corresponding PSFs. All size measurements are reported in Table~\ref{tab2}. 

We also compute the ellipticities of the isophotes corresponding to the radii listed above. Ellipticities are defined as $\epsilon \equiv 1-b/a$ where $a$ and $b$ are the semi-major and semi-minor axis, respectively, and are reported in Table~\ref{tab2} except for ellipticities at $R_{\rm eff}$ which are affected by the PSF and are thus not reported. The nebulae of radio-quiet quasars are nearly round, with $\epsilon_{5\sigma}=0.13\pm0.12$ (mean and standard deviation). 

Furthermore, we create continuum maps for all objects by summing up the spectrum over a 200-300\AA\ wavelength range where line features are absent and artifacts are minimal (typically either 4400--4700\AA\ or 5200--5400\AA), and perform differential photometry on these maps in the identical manner to that employed for \oiii\ line images. We list both the $R_{5\sigma}^{\rm cont}$ and the $R_{\rm eff}^{\rm cont}$ in Table~\ref{tab2}. Our sensitivity to the continuum observations is limited, since the continuum is extracted using a relatively narrow wavelength range; therefore, the continuum profiles shown in Figure \ref{fig:psf} with dashed lines are truncated at much smaller distances from the center than the profiles of the nebular emission. The continua are spatially resolved, with $R_{\rm eff}^{\rm cont}$ ranging from 1.3 to 2.3 times the effective radii of the corresponding PSFs, and are nearly round for the radio-quiet quasars, with $\epsilon_{5\sigma}^{\rm cont}=0.14\pm 0.05$. 

\begin{figure*}
\centering
\includegraphics[scale=0.6,angle=0]{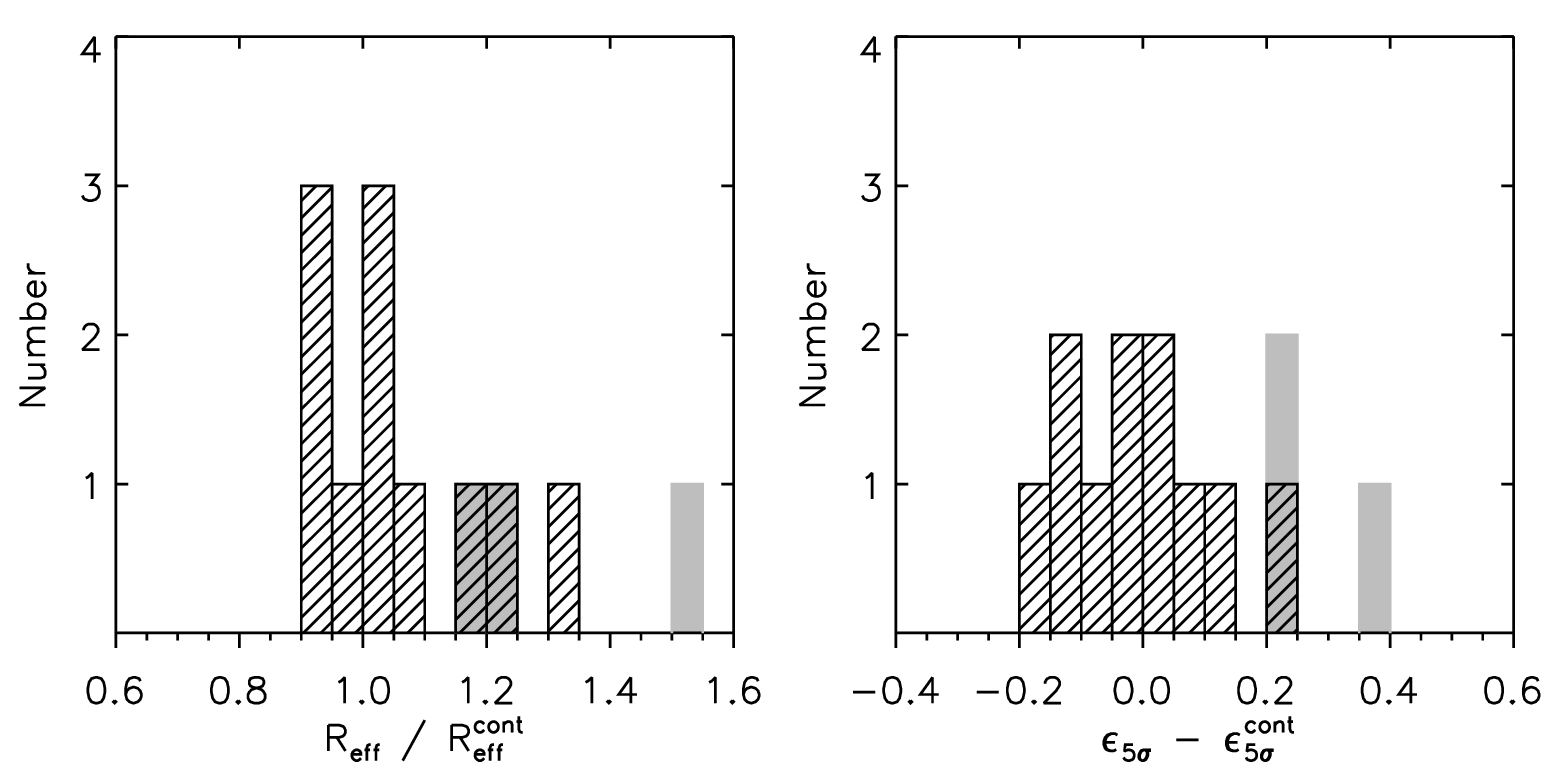}
\caption{Comparison of the extents and ellipticities between the [O {\sc iii}]$\lambda$5007
line and the continuum maps of our quasars. In the eleven radio-quiet 
objects (striped bars), the effective radii of the nebulae and the continua are similar, as are their ellipticities. The three radio-loud objects are represented by gray solid bars.}

\label{fig:ct_hist}
\end{figure*}

In Figure~\ref{fig:ct_hist} we show the distribution of relative sizes and relative ellipticities for the nebulae and the continua. For radio-quiet objects, the ratio $R_{\rm eff}/R_{\rm eff}^{\rm cont}$ ranges between 0.90 and 1.33, with the mean and standard deviation of $1.11\pm 0.18$. The continuum is likely due to the combination of the starlight from the host galaxy and the small amount of light from the obscured quasar that escapes along unobscured directions, scatters off the interstellar medium of the galaxy and reaches the observer. Since scattered light is polarized, polarimetry or spectropolarimetry can provide a way to determine the dominant direction of quasar illumination \citep{anto85, tran95, zaka05, zaka06}. The contribution of the scattered light to the total continuum emission ranges from 0 to 60\% \citep{liu09}. Thus the spatial profile of the continuum is determined by both the distribution of stars in the galaxy and the distribution of the interstellar medium, whereas the sizes of the emission line nebulae are determined by the distribution of the ionized gas.

The small ellipticities of the continuum emission suggest that the scattered light emission (which tends to show pronounced conical or biconical structure; \citealt{zaka06}) is either not an important contributor or is concentrated close to the center of the galaxy on scales smaller than the seeing. Of the sources presented here, spectropolarimetric observations are available for SDSS~J0842+3625. This object's continuum emission is polarized at an astonishing 20\% level, and there are no discernible stellar features in the deep optical spectrum we took of this source \citep{zaka05}. These data indicate that much, if not most, of the continuum is due to the scattered light. The position angle of the E-vector of polarization is 11$^{\circ}$ east of north and therefore the dominant illumination direction is 101$^{\circ}$ east of north, as indicated in Figure~\ref{fig:OIII} with dotted lines. Although some elongation of \oiii\ emission is vaguely visible in this direction, the continuum morphology shows no features. Therefore, even in this rather extreme source the scattered light emission is not spatially resolved by the ground-based observations, and the combination of \emph{HST} and polarimetric observations would be necessary to determine the relative importance of starlight and scattered light in our sample. 

Since the narrow-line emission and the continuum are produced by different components of the galaxy via different emission mechanisms, the close similarity of the effective sizes of the nebulae and the continua is rather surprising. The key question is whether the warm ionized gas observed via its line emission is co-spatial with stars or whether the gas is in fact being pushed out of the galaxy by the quasar into the intergalactic medium. Therefore, the proper comparison would be not between the effective radii of the nebular and continuum emission, but rather between the density profiles of the ionized gas and the stars. We hope to be able to obtain the density profiles of the ionized gas by extending the range of our photo-ionization models of the kind presented in Section \ref{sec:model}. To get accurate stellar light profiles would require \emph{HST}-quality data to separate out the scattered light contribution \citep{zaka05, zaka06}. 

\subsection{Size-Luminosity Relation}

\begin{figure*}
\centering
\includegraphics[scale=0.7,angle=0]{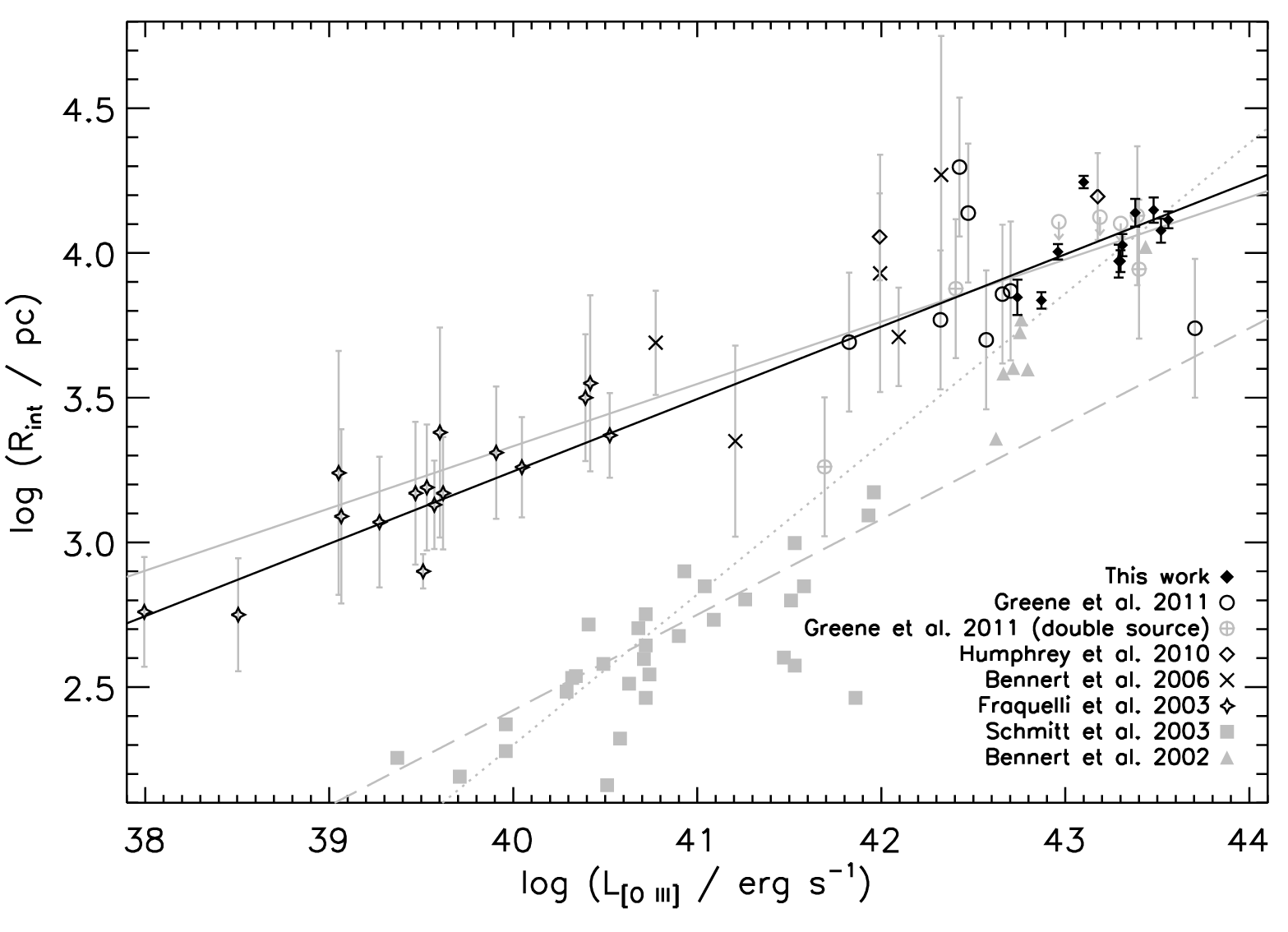}
\caption{Size-luminosity relation of the radio-quiet quasars in the \oiii\ 
emission line from our sample (black solid diamonds), added to the previous 
observations including IFU mapping \citep{hump10} and long-slit 
spectroscopy \citep{gree11} of type 2 quasars at $z$=0.1--0.4, and 
long-slit spectroscopy of nearby Seyfert 2 galaxies \citep{benn06,fraq03}. 
\emph{HST} narrow-band imaging campaigns \citep{schm03,benn02} are shown with 
gray symbols but are not used in any quantitative analyses, 
because they are shallower than IFU and long-slit spectroscopy 
by about 1 to 2 orders of magnitude \citep{gree11}.} 
\label{fig:sl}
\end{figure*}

We now investigate the relationship between the integrated \oiii\ luminosity and the size of the nebula -- the size/luminosity relationship shown in Figure~\ref{fig:sl}. In addition to our own data on the high-luminosity end of the diagram, we use several samples from the literature. These datasets include IFU maps of two type 2 quasars \citep{hump10}, long-slit spectroscopy of type 2 quasars at $z$=0.1--0.4 \citep{gree11}, and 
long-slit spectroscopy of nearby Seyfert 2 galaxies \citep{benn06,fraq03}. 
\emph{HST} narrow-band imaging campaigns \citep{schm03,benn02} are included in 
the figure (represented by gray symbols) but are not used in any 
quantitative measurements, because they are shallower than IFU and 
long-slit spectroscopy by 1--2 orders of magnitude and thus do not
reach down to the surface brightness levels we use in our analysis
\citep{gree11}. 

Except for the solid gray symbols, the radius reported in Figure~\ref{fig:sl} 
is $R_{\rm int}$, the isophotal radius at 
$10^{-15}/(1+z)^4$ erg s$^{-1}$ cm$^{-2}$ arcsec$^{-2}$
which corrects the sizes for cosmological dimming and is thus the measure of choice for samples that include objects at different redshifts.
For our data, we employ a conservative approach to estimate 
the uncertainty of $R_{\rm int}$. Taking into account that the SDSS 
calibration error is $\lesssim10\%$ (Section \ref{sec:calib}), and that the 
isophote at $R_{\rm int}$ corresponds to a signal-to-noise ratio of 3.5--10.7
among the radio-quiet objects, we find a 10-30\% uncertainty. This is an overestimate, because measuring an isophotal radius involves averaging
many pixels and the uncertainty is thus suppressed; but the pixels are 
actually not independent and the uncertainty is difficult to further 
quantify. We budget a further 30\% uncertainty to account for continuum subtraction and instrumental and pipeline errors, and combining all these uncertainties in quadrature leads to a final uncertainty of $\la 40\%$ in our surface brightness measurements. Since the surface brightness declines as a steep function of radius, we find that the uncertainty in our determined $R_{\rm int}$ are about
5--14\% for the eleven radio-quiet quasars.

We remeasure the radii of the targets in the four relevant studies \citep{fraq03, benn06, hump10, gree11} to determine 
$R_{\rm int}$ in a uniform manner to construct the size-luminosity relation. 
Among the six objects observed by \citet{hump10}, only two radio-quiet
quasars are resolved and found to be surrounded by ionized gas nebulae.
We assign a 50\% uncertainty for their radii, which is an arbitrary 
but conservative estimate from their IFU \oiii\ maps with 1\arcsec\ 
spaxels and 1.2\arcsec\--1.6\arcsec\ seeing. All the 15 quasars in the 
\citet{gree11} sample are measured using the 1-dimensional spatial 
profile of \oiii, and we adopt the uncertainty of about 50\% estimated by \citet{gree11} to account for the unknown elongation of these nebulae.

For the objects in the \citet{benn06} and \citet{fraq03} samples the 
cosmological dimming effect of surface brightness is not important 
because these Seyfert 2 galaxies are nearby (the most distant object 
CGCG 420-015 is at $z=0.0294$). Our limiting surface brightness level
corresponds to $10^{33.7}$ erg s$^{-1}$ pc$^{-2}$ in their units for 
all these targets. Both groups have provided the best-fit power laws 
of the \oiii\ surface brightness profiles, which we use to obtain 
$R_{\rm int}$ \citep[the \oiii\ profile of ESO 362--G008 is not 
available and we exclude this source from the sample of][]{benn06}. For 
the two targets included in both samples, NGC 1386 and NGC 5643, we 
choose the profiles given in \citet{benn06}, because these authors 
subtract the contribution of the star formation to the ionization 
of \oiii. We adopt the uncertainties estimated by \citet{fraq03} for
their objects, while for the five \citet{benn06} objects we propagate 
the uncertainties in the best-fit power indices to $R_{\rm int}$. 

The uniformly defined sizes from previous investigations and the addition 
of our high-quality size measurements at high luminosity are combined into a single size-luminosity relationship in Figure \ref{fig:sl}. The $R_{\rm int}$ sizes are strongly correlated with \oiii\ luminosity, and we fit the relationship with a power law. We perform a linear regression on the well-measured logarithmic data points, 
with upper limits and double continuum sources from \citet{gree11} excluded. 
We use linear regression algorithm {\sl linmix\_err} from the 
IDL Astronomy User's Library\footnote{http://idlastro.gsfc.nasa.gov/},
which applies a Bayesian approach using Markov chain Monte Carlo to
calculate the posterior probabilities as described in \citet{kell07}. 
By fitting a Gaussian profile to the resultant posterior distribution of 
the slope and intercept which are both close to normal 
distributions, we take advantage of Bayesian methods to reliably 
calculate the uncertainties. Assuming a 15\% uncertainty for all 
$L_{\rm [O~{\scriptscriptstyle III}]}$ measurements, we find a best-fit 
linear relation

\begin{center}
log($R_{\rm int}$/pc)=($0.250\pm0.018$) 
log($L_{\rm [O~{\scriptscriptstyle III}]}/10^{42}$ erg s$^{-1}$)+($3.746\pm0.028$).
\end{center}

The best-fit parameters are consistent within 1$\sigma$ with those found 
by \citet{gree11}. In particular, we confirm the shallow slope they find 
($0.22\pm0.04$). We caution that
the shallow slope hinges in part on the data points at the faint end from \citet{fraq03}, 
which are based on extrapolations from their power-law fit to the \oiii\ profile 
rather than real measurements \citep[see also discussion in][]{huse12}.

We detect extended \oiii\ emission in every case, whereas \citet{hump10} conducted IFU observations of type 2 quasars drawn from the same parent sample \citep{zaka03, reye08} and resolved two out of six objects. The most likely reasons for this difference are the smaller luminosity of the objects selected by \citet{hump10} (which implies smaller expected sizes, as per our size-luminosity diagram) and the 2--3 times better seeing of our observations. 

\subsection{Emission line ratios}
\label{sec:line_ratio}

\begin{figure*}
\centering
    \includegraphics[angle=0,origin=c,scale=.59]{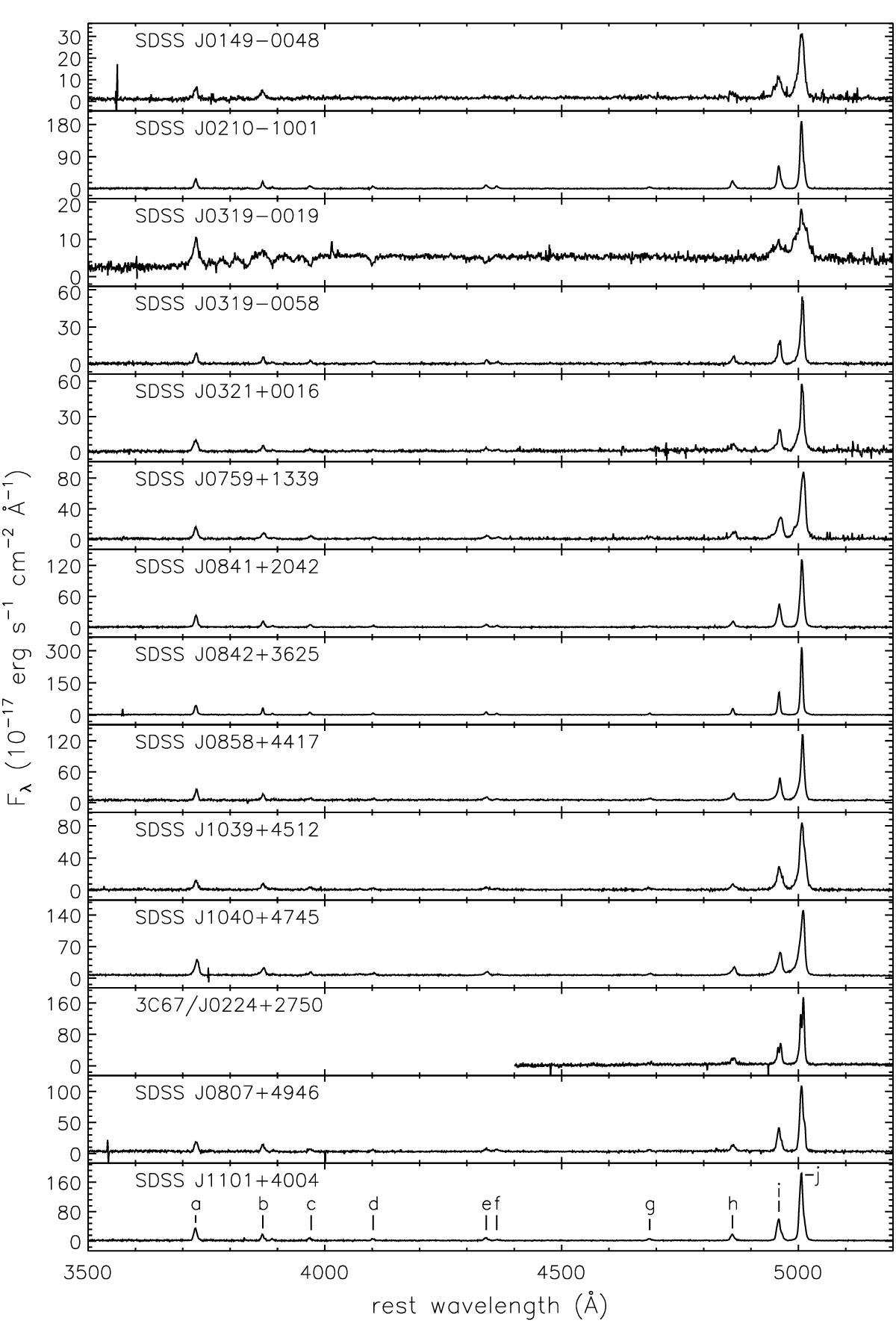}%
    \includegraphics[angle=0,origin=c,scale=.59]{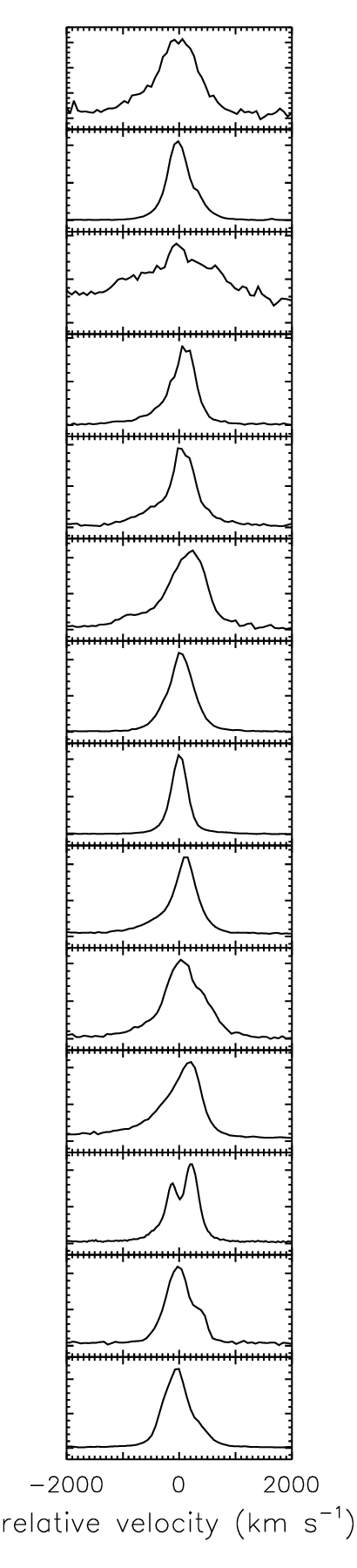}
\caption{Spectra of our sample quasars, taken from the SDSS DR7 except
for 3C67 whose spectrum is obtained by collapsing our Gemini IFU 
datacube along spatial dimensions. The original SDSS spectra have much
broader wavelength coverage than the GMOS spectra examined in this
paper, but do not contain any spatial information. 
The marked emission lines are: 
($a$) [O {\sc ii}] 3727\AA; ($b$) [Ne {\sc iii}] 3869\AA; 
($c$) [Ne {\sc iii}] 3871\AA; ($d$) H$\delta$ 4102\AA; 
($e$) H$\gamma$ 4341\AA; ($f$) \oiii\ 4363\AA; ($g$) He {\sc ii} 4686\AA;
($h$) H$\beta$ 4861\AA; ($i$) \oiii\ 4959\AA; ($j$) \oiii\ 5007\AA. 
Right panels show the velocity substructures in the [O {\sc iii}] 5007\AA\ 
line in the rest frame of the sources as determine using the redshifts
listed in Table~\ref{tab1}. The GMOS spectra analyzed in this paper
typically cover He {\sc ii}, H$\beta$ and \oiii.}
\label{fig:sdss}
\end{figure*}

Among the various optical line ratios, \oiii/H$\beta$ is one of the best 
diagnostics of the ionization mechanism \citep{bald81,veil87,oste06}.
In order to create the intensity maps of the H$\beta$ emission line, we first need to evaluate the contribution of the host galaxy. Unlike the case of \oiii\ where a local linear continuum fit is sufficient, the measurement of H$\beta$ may be biased by the absorption lines in the stellar photospheres of the host galaxy. We take the fiber spectrum of 
each object from the SDSS database and decompose the continuum into a
linear sum of the principle components derived from pure absorption-line galaxies \citep{hao05}. We subtract the best-fit host galaxy continuum from each SDSS spectrum and find
that the H$\beta$ flux changes by no more than a few percent, with the
exception of SDSS J0319-0019 (Figure~\ref{fig:sdss}). This object shows a strong post-starburst continuum \citep{goto03}, and subtraction of this component leads to a detection of H$\beta$ emission with a $S/N\sim 2.5$ at the peak of the line (which is otherwise undetected in the total spectrum). We attempted to apply this strategy to the IFU data of this target in a spatially resolved manner, but could not reliably recover the extended H$\beta$ emission, because the baseline 
determination is hindered by the chip gap in vicinity of the H$\beta$ line. Therefore, we
do not perform spatially resolved analysis of \oiii/H$\beta$ in this one target. In the remaining objects, we create the intensity maps of the H$\beta$ line emission from our sample 
objects in an identical manner as for the [O {\sc iii}] line. 

To avoid oversampling within the PSF and to increase the $S/N$ ratio, we rebin 
the intensity maps so that the pixel size matches the full width at 1$\sigma$ 
of the Gaussian profile (i.e., pixel scale = 2$\sigma$) that represents the 
corresponding observational PSF. In Figure~\ref{fig:o3hb}, the [O {\sc iii}]/H$\beta$ 
line ratio is plotted against the [O {\sc iii}] line surface brightness 
($\Sigma_{\rm [O~{\scriptscriptstyle III}]}$) 
and the corresponding isophotal radius (semi-major axis from the elliptical 
differential photometry of the [O {\sc iii}]$\lambda$5007 map, see 
Section~\ref{sec:psf}). In this spatially resolved pixel-by-pixel analysis, 
we have only considered pixels with $S/N\geqslant3$ 
in both lines. Although the three radio-loud and radio-intermediate quasars 
are included, their isophotal morphology is not as well represented by ellipses 
as radio-quiet objects, therefore the 
$\Sigma_{\rm [O~{\scriptscriptstyle III}]}$-to-radius conversion introduces
larger uncertainties.

Figure~\ref{fig:o3hb} reveals a universal behavior of the [O {\sc iii}]/H$\beta$ 
line ratio in our entire sample of radio-quiet quasars: the ratio persists at
a constant value of roughly 10 in the central regions, until it reaches a ``break'' 
isophotal radius ranging from $\sim$4.1 to 11 kpc where it starts to decrease.
In Section~\ref{sec:model} we further discuss this figure and present our 
qualitative interpretation of this behavior.

In addition to H$\beta$ and \oiii\ we detect several weaker emission
lines. Because of their lower total fluxes, they are typically
detected out to smaller distances than H$\beta$ (and of course smaller
yet than \oiii). One of these, \heii~$\lambda$4686\AA, turns out to be
diagnostic of physical conditions in the nebula. We describe the
underlying physics and present the \heii~$\lambda$4686\AA\
measurements in Section \ref{sec:model}. 

The wavelength coverage of the GMOS spectra is rather limited, and H$\gamma$ is covered only in a couple of objects in the region of poor filter transmission. The typical reddening of the narrow-line regions of type 2 quasars was estimated by \citep{reye08} from the spatially integrated SDSS spectra to be $A_V=1.0-2.6$ mag based on the observed H$\alpha$/H$\beta$ and H$\beta$/H$\gamma$ ratios. The only two line ratios we are discussing in the present paper are \oiii/H$\beta$ and \heii/H$\beta$. Because all these lines are close together in wavelength, de-reddening would result in a small correction to these ratios of $\la 0.06$ dex. 

\begin{figure*}
\centering
    \includegraphics[totalheight=3cm,angle=0,origin=c,scale=.92]{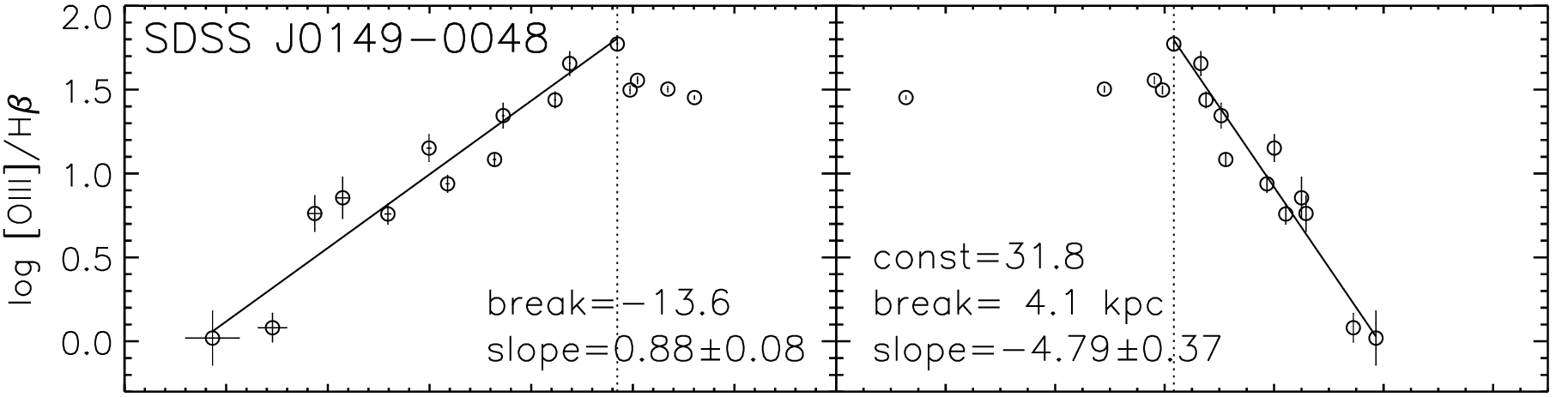}
    \includegraphics[totalheight=3cm,angle=0,origin=c,scale=.92]{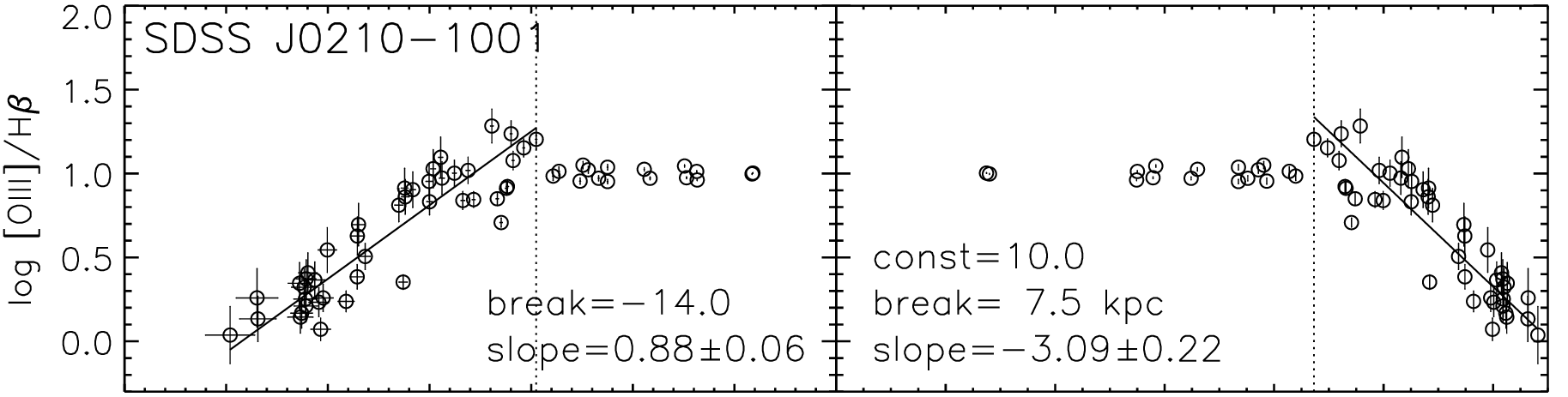}
    \includegraphics[totalheight=3cm,angle=0,origin=c,scale=.92]{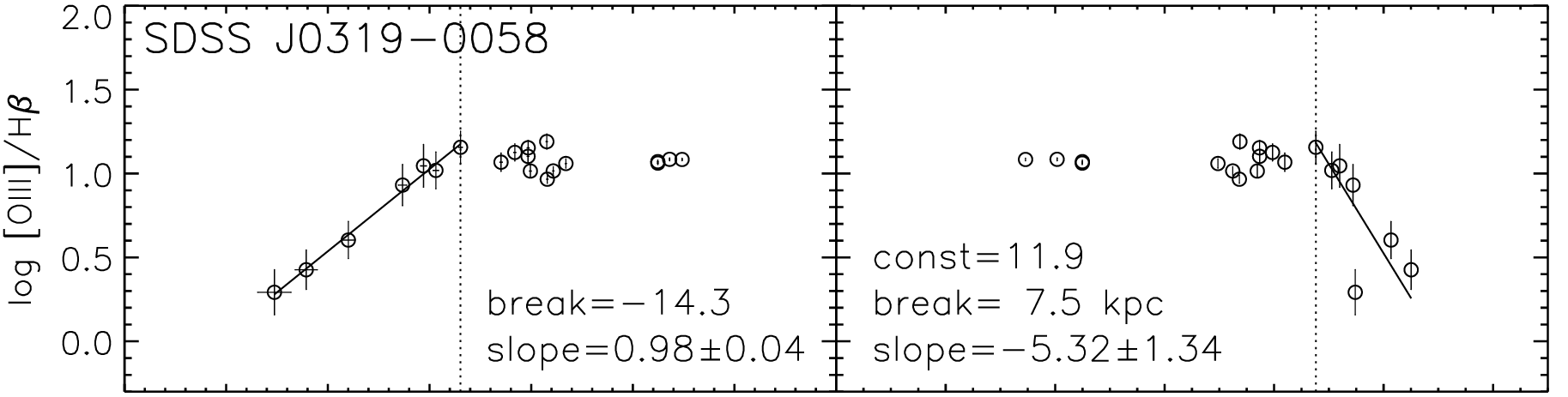}
    \includegraphics[totalheight=3cm,angle=0,origin=c,scale=.92]{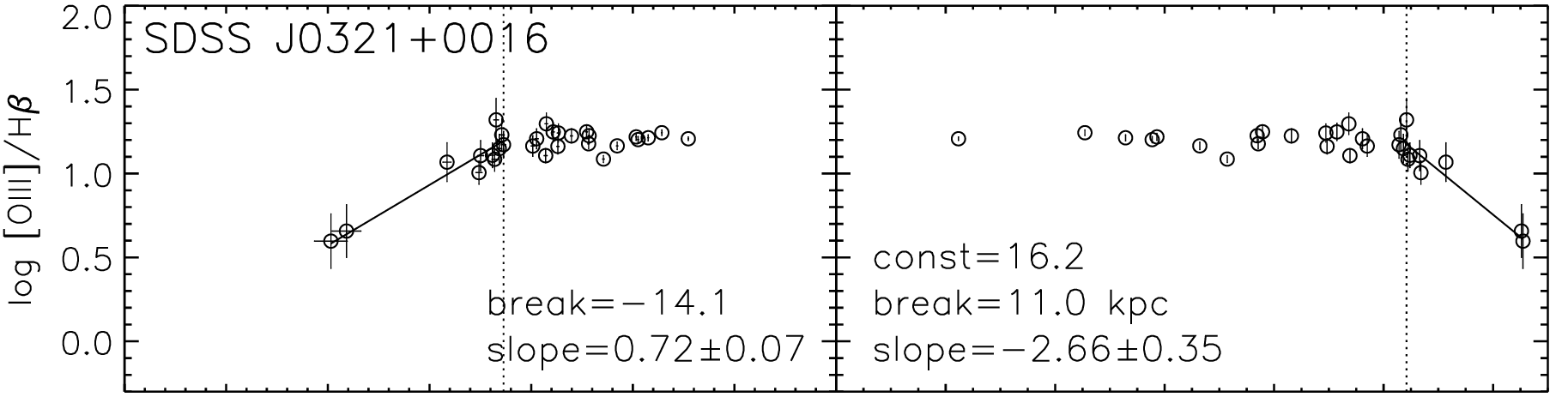}
    \includegraphics[totalheight=3cm,angle=0,origin=c,scale=.92]{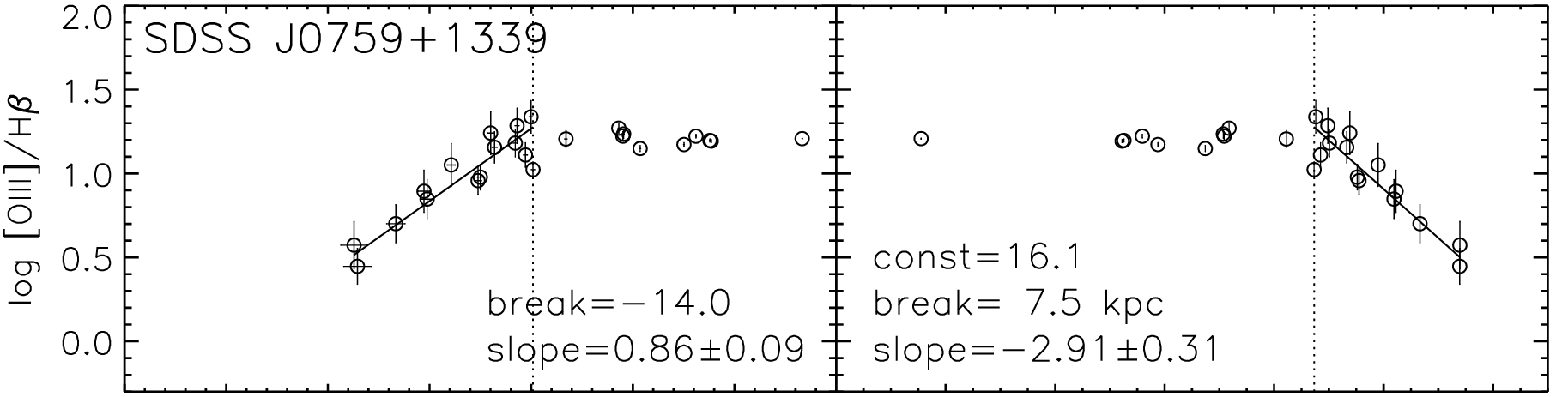}
    \includegraphics[totalheight=3cm,angle=0,origin=c,scale=.92]{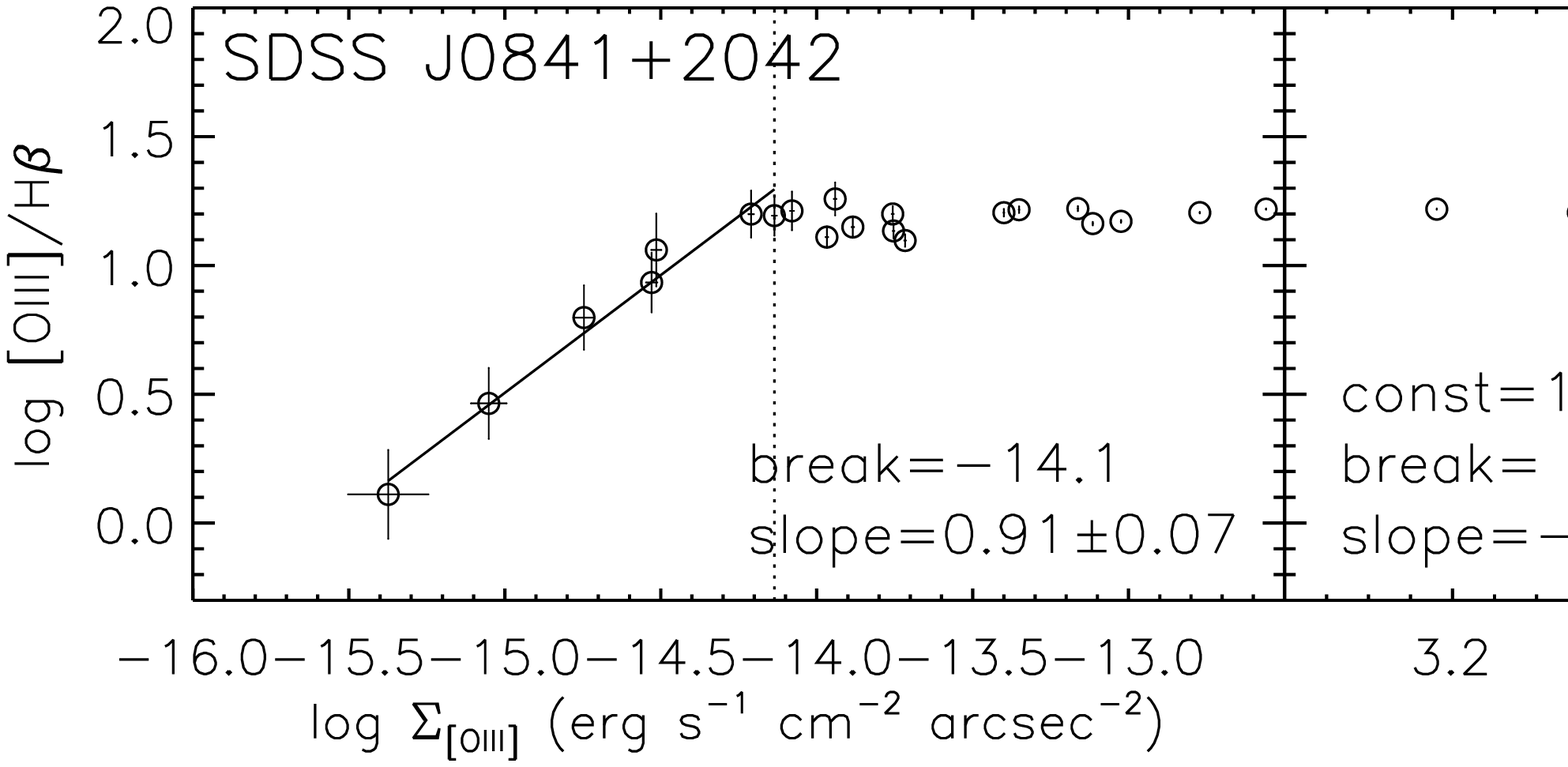}\\
\vspace{7mm}
\caption{[O {\sc iii}]/H$\beta$ as a function of [O {\sc iii}] surface brightness and 
isophotal radius (semi-major axis). For each object, we list the \oiii/H$\beta$ ratio on the 
plateau (labeled ``const''), the \oiii\ surface brightness and radius where the break occurs, and the 
best-fit power-law slope of the non-plateau regimes for each object are marked.
The breaks are depicted by dotted lines.}
\label{fig:o3hb}
\end{figure*}

\begin{figure*}
\centering
    \includegraphics[totalheight=3cm,angle=0,origin=c,scale=.92]{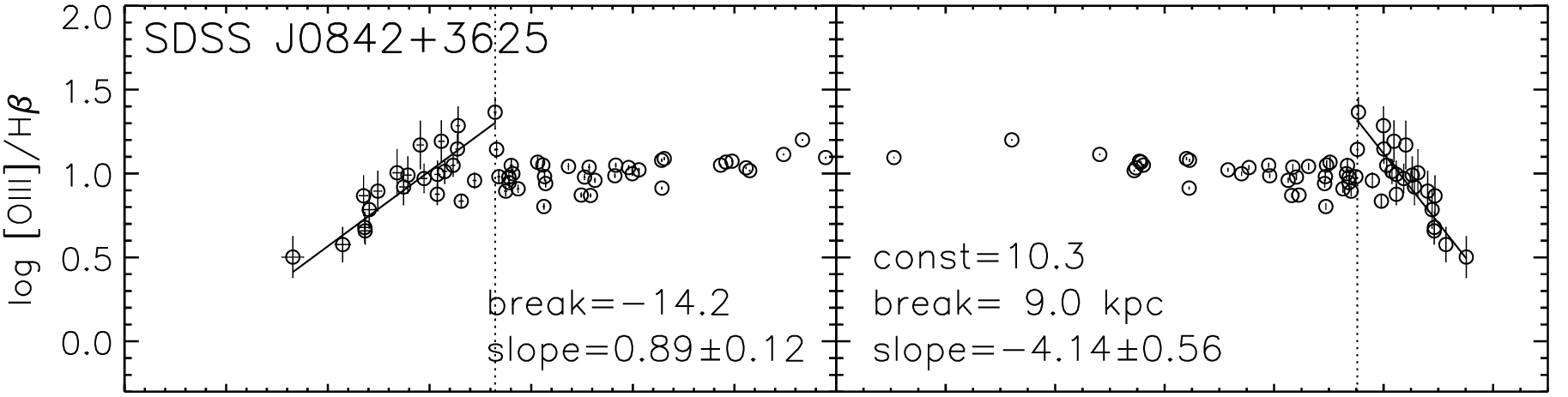}
    \includegraphics[totalheight=3cm,angle=0,origin=c,scale=.92]{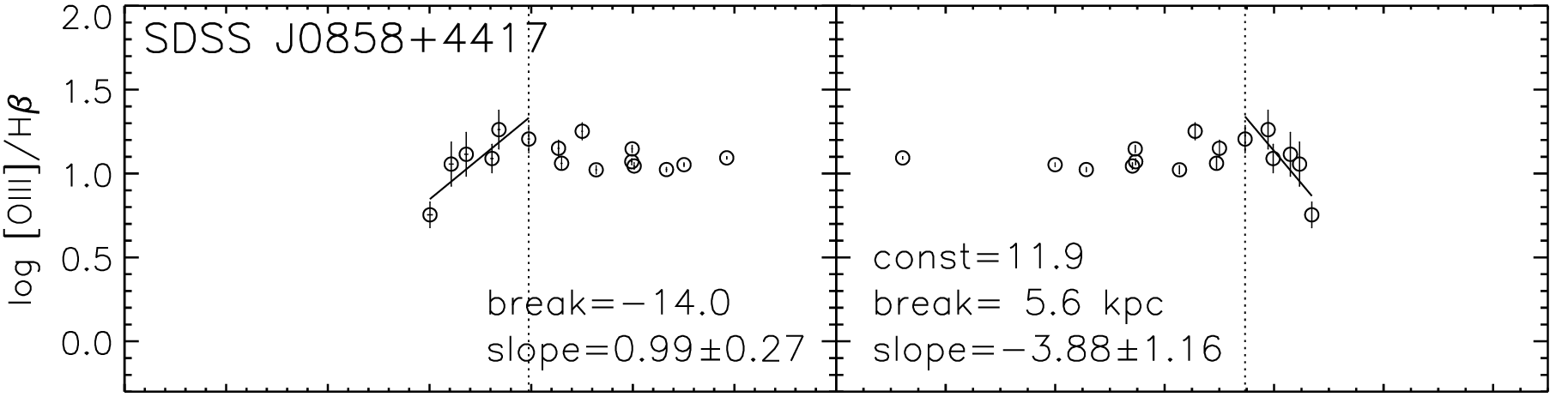}
    \includegraphics[totalheight=3cm,angle=0,origin=c,scale=.92]{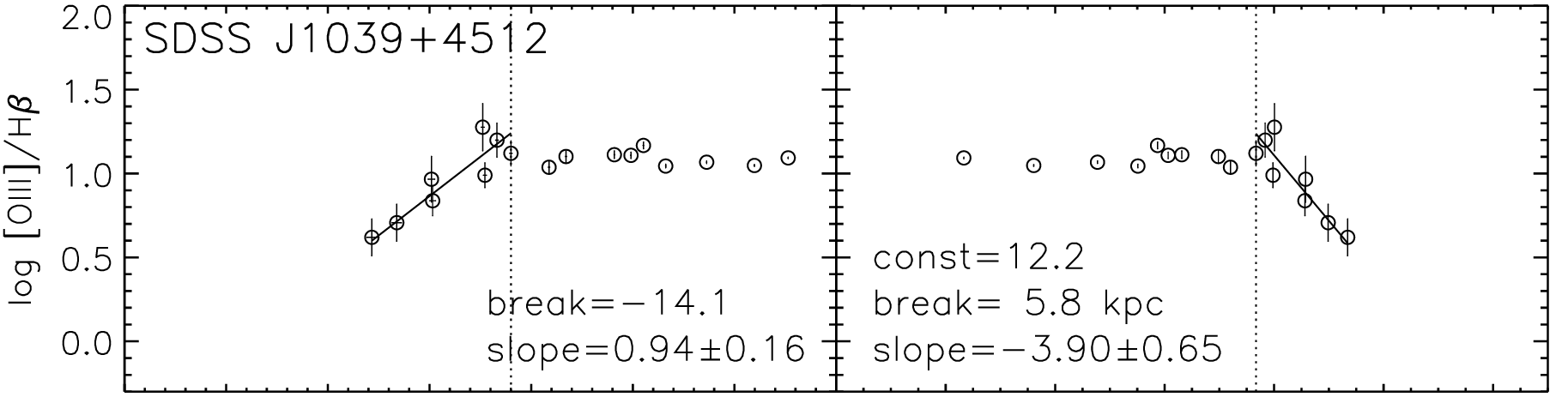}
    \includegraphics[totalheight=3cm,angle=0,origin=c,scale=.92]{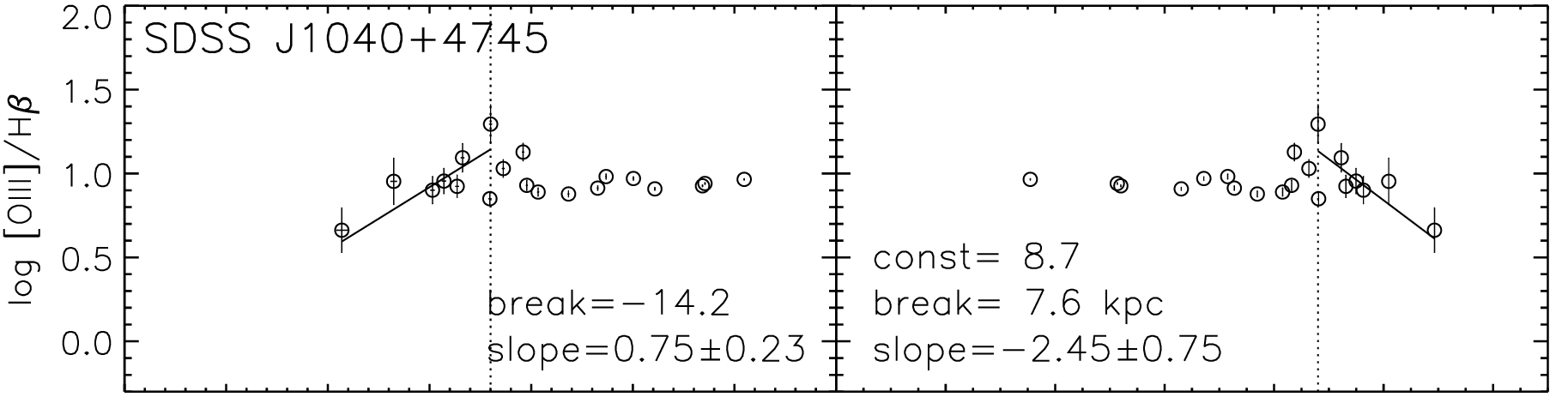}
    \includegraphics[totalheight=3cm,angle=0,origin=c,scale=.92]{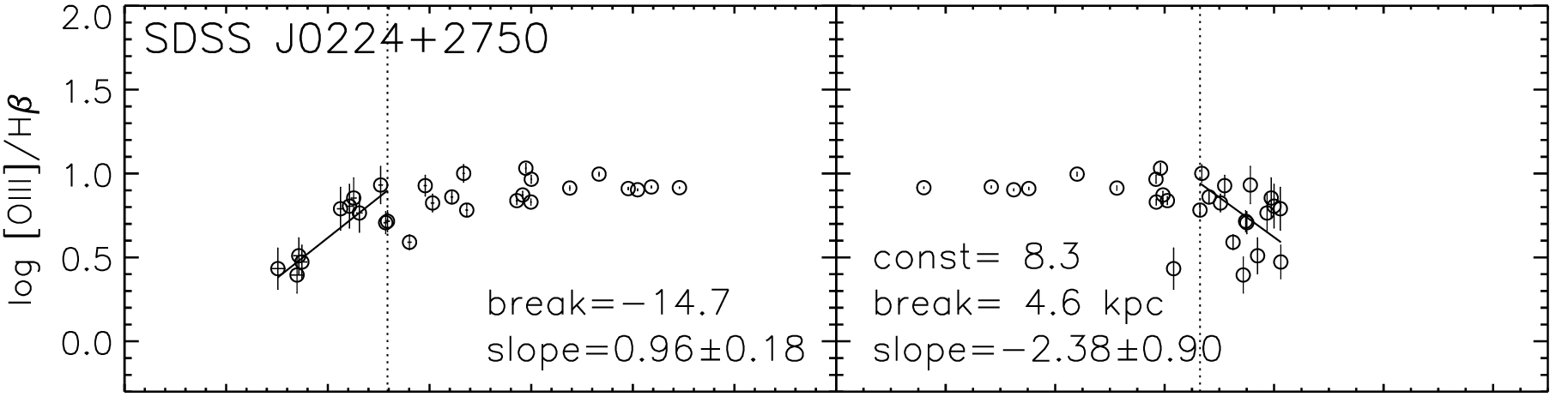}
    \includegraphics[totalheight=3cm,angle=0,origin=c,scale=.92]{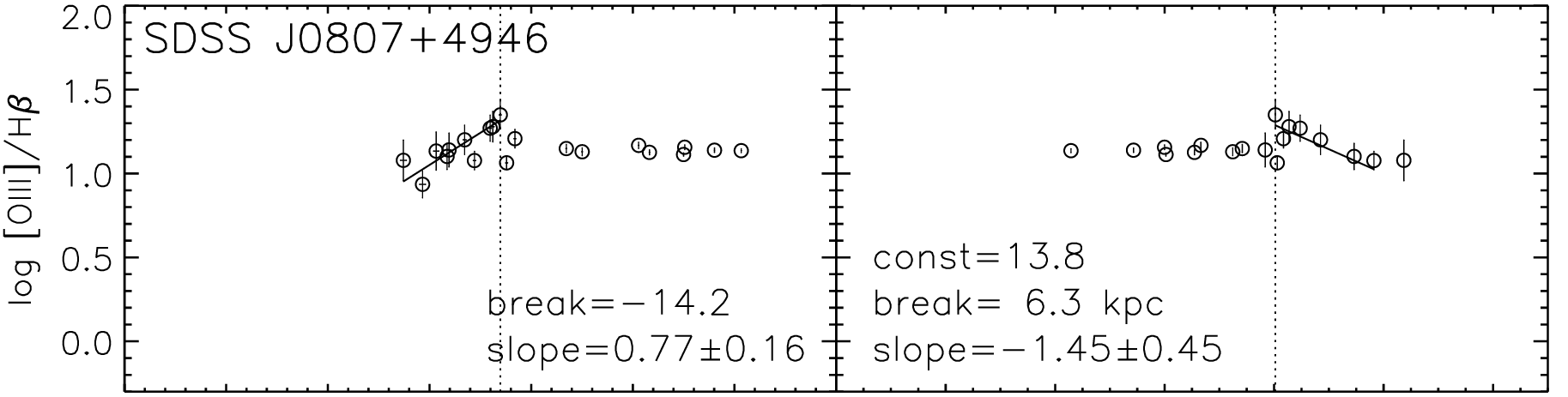}
    \includegraphics[totalheight=3cm,angle=0,origin=c,scale=.92]{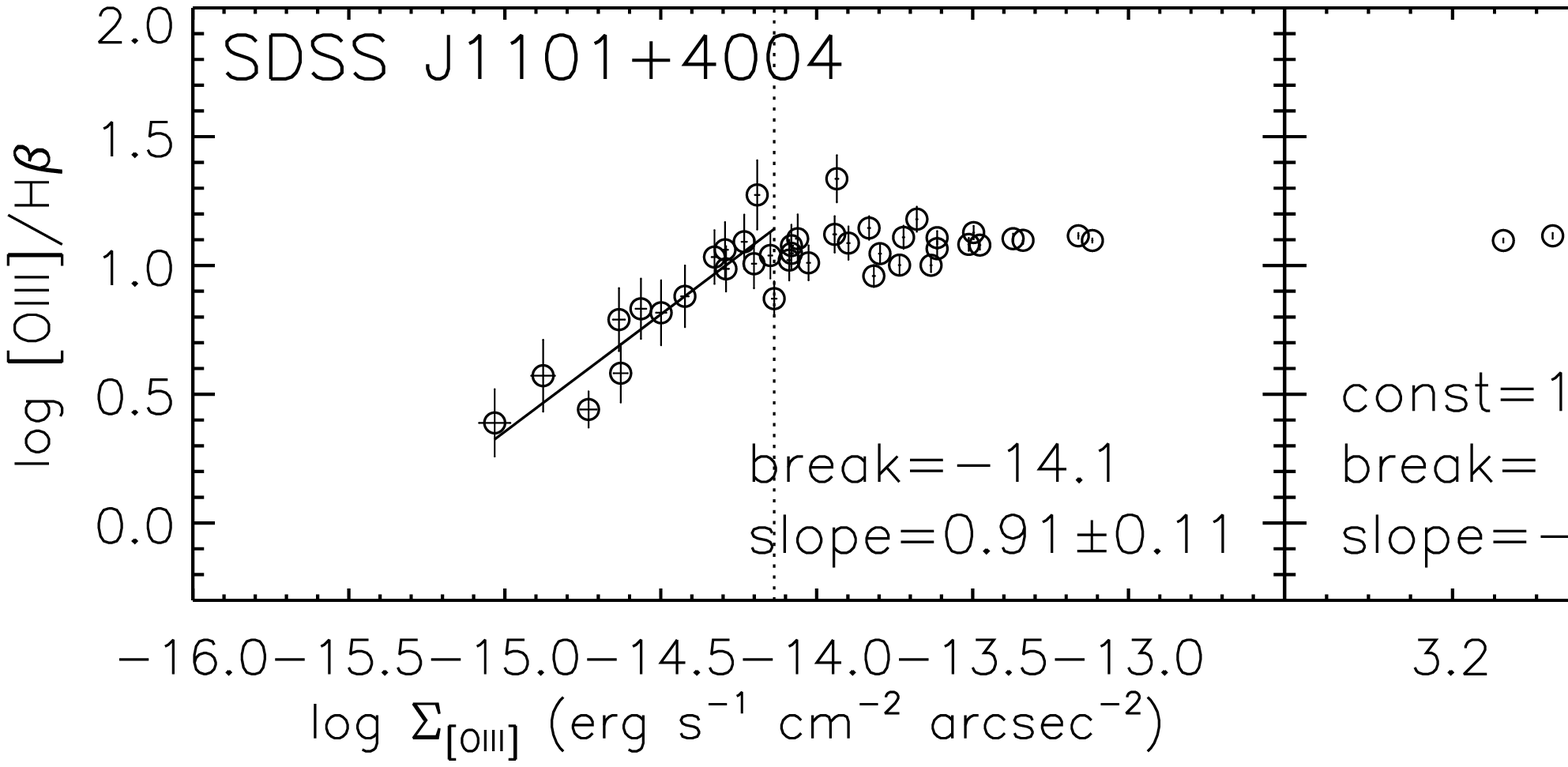}\\
\vspace{1cm}
\caption{Continued.}
\end{figure*}
\clearpage

\section{Discussion}
\label{sec:discussion}

\subsection{Comparison with radio-loud objects} 
\label{sec:RL}

We find extended ($\gtrsim$ 20 kpc-scale) ionized gas nebula in every highly luminous 
radio-quiet obscured quasar that we have examined at $z\sim0.5$. This finding is in 
contrast to many previous studies, beginning with \citet{stoc87}, which found few if any radio-quiet objects with extended line emission; rather, it was the radio-loud quasars that routinely displayed powerful extended emission-line regions. Since we are now able to extend previous studies to a higher quasar luminosity, particularly for obscured quasars, we revisit the comparison between the properties of the nebulae of radio-loud and radio-quiet objects. 

Jets are frequently implicated in strong interactions with galactic interstellar matter \citep{oost00}, in exciting the gas through shocks \citep{ceci01} and in producing outflows of gas from the host galaxy, often on scales of several kpc \citep{baum90, baum92, mcca93, morg05, nesv06}. At high redshifts, the emission-line regions can be extended over tens of kpc, contain significant amounts of mass and are typically well-aligned with the large-scale jet emission \citep{nesv06, nesv08}. At low redshifts, there is no particularly strong alignment between the jet and the emission-line gas \citep{priv08}, perhaps with marginal evidence for alignment in younger radio sources still confined to their host galaxy. The lack of alignment could have many explanations -- for example, the pattern of quasar illumination of the interstellar medium may be different from the jet direction, or inhomogeneities in the gas could lead to nebulae expanding in the directions of lowest density. No obvious alignment is seen in the three radio-loud sources we observed as part of our campaign (bottom of Fig. \ref{fig:OIII}). Of the four extended emission features seen in our three radio-loud sources, only one feature in SDSS~J0807+4946 lies along its radio axis. 

For comparison with our objects, an ideal radio-loud sample would contain type 2 sources (i.e., radio galaxies) with similar redshifts and \oiii\ luminosities observed to similar depth. Narrow-band imaging data from space \citep[e.g., for 3C objects,][]{trem09} are therefore not a good option because they are shallower than our IFU data by 1--2 orders of magnitude. A close-to-optimal sample was studied by \citet{fu09} who consider eight radio-loud sources. Their objects are type 1 radio-loud quasars which are a bit closer than ours ($z=0.16-0.37$) and have slightly lower luminosities ($L_{\rm [O~{\scriptscriptstyle III}]}=10^{42\mhyphen43}$ erg s$^{-1}$), but their IFU data is very well matched to our observations. 

The most striking difference between our radio-quiet objects and the radio-loud sources studied by \citet{fu09} is that the objects from the latter sample show much more complex morphologies in their \oiii\ emission. The nebulae studied by \citet{fu09} often consist of several kinematically and/or morphologically separate clumps and blobs, and even when the emission is connected into a single nebula it is not a linear structure that we could fit with an ellipse for comparison with our observations. \citet{fu09} suggest that the blobs and clumps could be due to quasar-illuminated merger debris or the remnants of a quasar-driven wind.

In view of the limited availability of comparison radio-loud objects at low redshifts, we consider high-redshift samples as well. We construct the radio-loud comparison sample by combining the three radio-loud quasars in our own Gemini campaign (Fig. \ref{fig:OIII}, bottom three) with three powerful radio galaxies MRC 0316-257, MRC 0406-244 and TXS 0828+193. These three objects, located at $z\sim$2--3, were observed with SINFONI in the IFU mode on the VLT by \citet{nesv08}. The isophotal radii $R_{\rm int}$ (semi-major axes) and ellipticities $\epsilon_{\rm int}$ are measured the same way as for radio-quiet quasars at the \oiii\ surface brightness of 10$^{-15}$ erg s$^{-1}$ cm$^{-2}$ arcsec$^{-2}$ (i.e., corrected for cosmological dimming). These results, along with some basic information about the three high-redshift objects, are listed in Table~\ref{tab3}.

\begin{figure*}
\centering
\includegraphics[scale=0.8,angle=0]{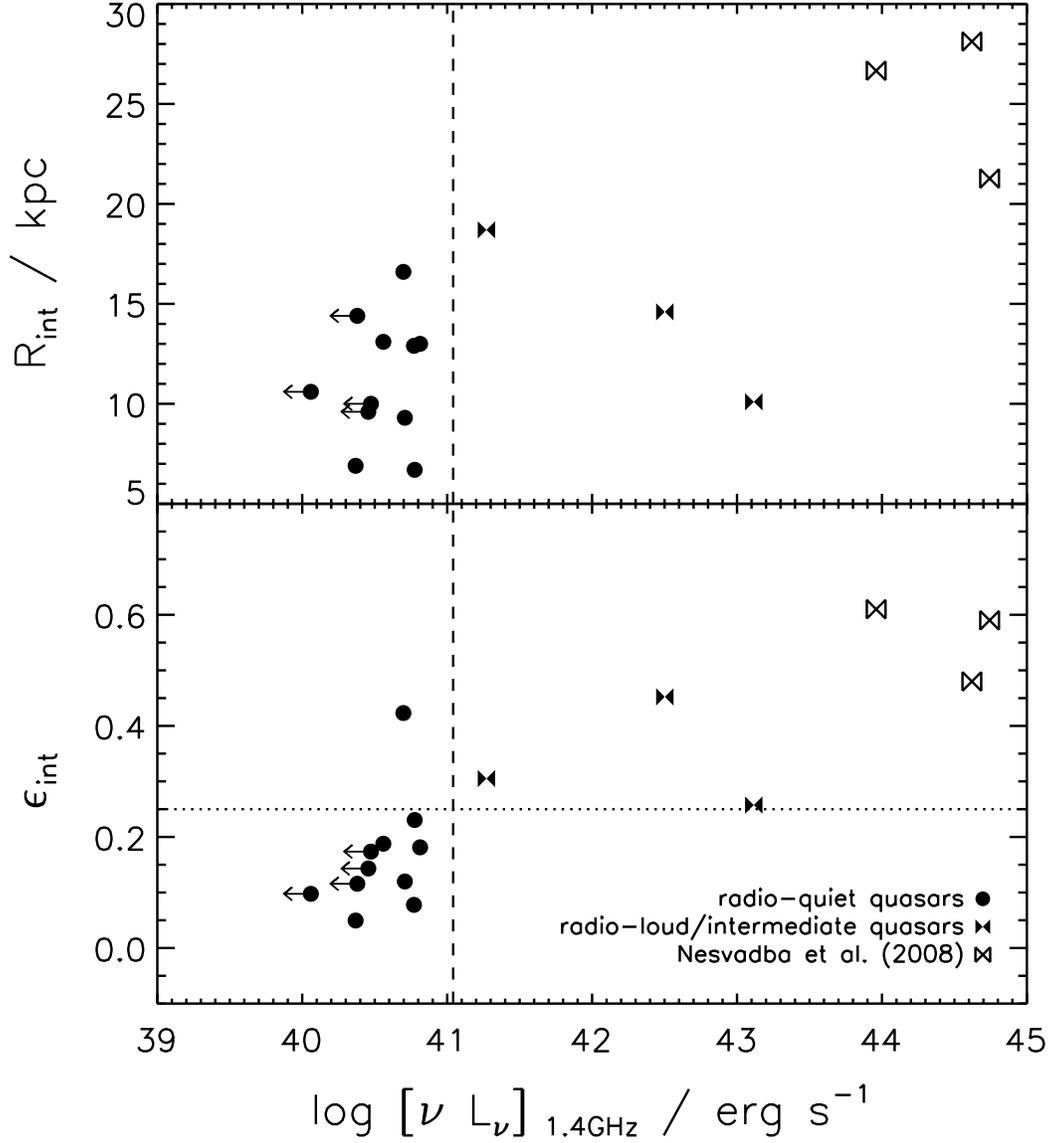}
\caption{Sizes and ellipticities ($R_{\rm int}$ and $\epsilon_{\rm int}$, 
see Section~\ref{sec:size} for their definition) of the \oiii\ nebulae around 11 radio-quiet 
quasars from our observations, compared to those of the six radio-loud objects (Table~\ref{tab3}). The comparison sample of six contains three low-redshift radio galaxies from our own observations and three high-redshift radio galaxies from \citet{nesv08}.} 
\label{fig:comp}
\end{figure*}

In Figure~\ref{fig:comp}, we show the radii $R_{\rm int}$ and ellipticities $\epsilon_{\rm int}$ of the eleven radio-quiet and the six radio-loud quasars plotted against their radio luminosity at 1.4 GHz. The 1.4 GHz fluxes are taken from the FIRST survey for SDSS objects and the NRAO/VLA Sky Survey \citep[NVSS,][]{cond98} for the rest and are then used to calculate the rest-frame-corrected monochromatic luminosities at 1.4 GHz \citep{zaka04}. In the case of non-detections in the FIRST survey, we report the upper limits for a point source \citep[5$\sigma$+0.25 mJy,][]{coll05}. 

Separated by the vertical dotted line at $\nu L_{\nu}[1.4{\rm GHz}]=1.1\times 10^{41}$ erg s$^{-1}$ (our adopted radio-quiet/radio-loud criterion discussed in Section~\ref{sec:selection}), the nebulae of the two populations of quasars demonstrate an evident difference in their ellipticities. All the radio-loud objects have elongated narrow-line nebulae with $\epsilon_{\rm int}>0.25$, while ten out of eleven radio-quiet quasars have $\epsilon_{\rm int}<0.25$, with SDSS J0321+0016 being the only exception. In terms of their respective mean value and standard deviation, the ellipticities of radio-quiet quasars have $\epsilon_{\rm int}=0.16\pm0.10$, while their radio-loud peers have $\epsilon_{\rm int}=0.45\pm0.14$. The most striking morphological difference between the ionized gas nebulae around radio-loud and radio-quiet objects is that the radio-quiet ones are significantly rounder. In contrast, the \oiii\ line emission from all the three objects in \citet{nesv08} are remarkably elongated along the direction of their radio jets. 

The semi-major axes of radio-loud nebulae ($R_{\rm int}=19.9\pm 6.9$ kpc) are larger than those of radio-quiet objects ($11.1\pm 3.0$ kpc), although the difference in sizes is not as dramatic as it is for ellipticities and there is some overlap in the size distributions. It is perhaps the larger sizes and larger ellipticities of the nebulae in radio-loud objects that made them more easily detectable in previous studies. 

Our radio-loud / radio-quiet criterion is set at $\nu L_{\nu}[1.4~{\rm GHz}]=1.1\times 10^{41}$ erg s$^{-1}$, which is a negligible fraction of the bolometric luminosity, but could correspond to a significant mechanical energy of the jet. For example, applying equations of \citet{cava10} we find that an $L_{\rm mech}=5\times 10^{44}$ erg s$^{-1}$ jet could be hiding behind this rather small radio luminosity. It is precisely the dramatic difference in morphology and ellipticity of the ionized gas nebulae that occurs at roughly this radio luminosity that leads us to conclude that jets are not an important contributor to supplying and exciting ionized gas in our eleven radio-quiet sources. 

\subsection{Comparison with Unobscured Quasars} 
\label{sec:T1}

Another natural comparison is between our sample and unobscured radio-quiet
quasars at similar luminosities and redshifts. Among the existing studies,
the sample of \citet{huse08} and \citep{huse12} is fairly well matched to ours 
in luminosity, although their sources are at slightly lower redshifts ($z<0.3$). 
At similar surface brightness limits, those authors do not find the spatially extended round nebulae that we see in our radio-quiet sample of type 2 quasars. They find extended \oiii\ emission in 61\% of the objects in their sample (the rest are non-detections). Compared to our objects, their nebulae appear elongated and/or irregular
(with ellipticities range from 0.15 to 0.64), and there is a trend for targets with high Fe {\sc ii} equivalent width (the doublet  $\lambda\lambda$4924,5018 plus the complex at 5100--5405\AA) and low H$\beta$ FWHM to have smaller nebulae. This last issue is not testable with our data because the Fe {\sc ii} emitting regions are obscured in our targets. These authors also find that the extended nebulae are aligned with the radio jets even when the jets are weak, and that the size of the nebulae increases for increasing radio luminosity.

One of the possible explanations for the difference between the detection rate of nebulae in our sample and in that by \citet{huse08} is that by selecting sources with high \oiii\ equivalent width (as is the case in our sample) we are 
preferentially selecting those with high covering factor in O$^{2+}$ 
\citep{bask05o3,ludw09}. Furthermore, despite the careful subtraction of the bright PSF (the quasar itself), faint emission on the scales of the host galaxy may be difficult to see in type 1 quasars, and thus the detected nebulae may be biased toward those that are larger and more elongated. Finally, geometric effects of quasar illumination may play a role: since type 2 quasars are obscured along the line of sight, the illumination of the gas may be happening on average closer to the plane of the sky than it does in type 1 quasars, thus producing apparently larger nebulae. 

Alternatively, the difference in the morphology of ionized gas around type 1 and type 2 quasars may reflect an actual evolutionary difference between the two types. Theoretical models of galaxy formation have long postulated an evolutionary scenario in which galaxy mergers induce both star formation and nuclear activity, triggering a transition from an obscured accretion and star formation stage (more likely to be seen as a type 2 source) to an unobscured phase as a type 1 quasar \citep[e.g.,][]{sand88}. A variety of observational studies support a scenario of this type, finding for example that host galaxies of type 2 active galactic nuclei exhibit more star formation than do the hosts of type 1's \citep{ho05,lacy07,zaka08}. Thus it is plausible that obscured quasars are more likely to be seen with extended ionized emission because they are in fact associated with a phase in which gas is being expelled out from the galaxy \citep{hopk06}.

\subsection{Comparison with Seyfert galaxies and the origin of gas}
\label{sec:sy2}

\citet{schm03} compiled and analyzed \emph{HST} narrow-band imaging data on \oiii\ nebulae in a large sample of Seyfert 1 and Seyfert 2 galaxies. The typical r.m.s. surface brightness limit of their observations is $1.2\times 10^{-15}$ erg s$^{-1}$ cm$^{-2}$ arcsec$^{-2}$, so these data probe the bright central regions of the targets on scales of a few hundred parsecs at typical resolutions of $\sim$50 pc. The median \oiii\ luminosity of the Seyfert galaxies in this sample is 2.5 dex smaller than the median \oiii\ luminosities of the quasars in our sample. 

Taking just the 38 type 2 Seyfert galaxies from this sample, we find that the ellipticities of their nebulae are distributed between 0.06 and 0.75, with the mean of 0.42. Many objects show pronounced biconical structures in their \oiii\ emission lines. Furthermore, we find 10 objects with position angles of the \oiii\ nebulae 
$\theta_{\rm [O~{\scriptscriptstyle III}]}$ measured by \citet{schm03} for which polarization position angles $\theta_{\rm pol}$ are available in the literature (as referenced in \citealt{schm03}). The average difference between the two sets of angles is $\langle|\Delta\theta|\rangle=97^{\circ}\pm6^{\circ}$. This value is in strong disagreement with that expected if $\theta_{\rm [O~{\scriptscriptstyle III}]}$ and 
$\theta_{\rm pol}$ were each independently drawn from a uniform distribution between 
$0^{\circ}$ and $180^{\circ}$, in which case the probability distribution of 
$|\Delta\theta|$ would be $p(\Delta\theta)\propto (1-|\Delta\theta|/180^{\circ})$ 
and would be peaked at small angles. 

As was already discussed in Section \ref{sec:size}, the polarization is produced when light from the hidden nucleus is scattered off of the extended material, and the illumination direction (as projected on the plane of the sky) is orthogonal to the measured polarization angle. Thus, in Seyfert 2 galaxies \oiii\ nebulae are not only elongated, but are closely aligned with the illumination direction. Both the scattered light and the \oiii\ emission are produced along the directions of clear view toward the nucleus, and in the classical model of toroidal obscuration \citep{anto85} one expects to see ionization bicones -- as indeed observed in the sample of \citet{schm03}.

In our sample, the observed morphologies of the nebulae are determined both by the gas distribution in and around the host galaxy and by the illumination pattern of the quasar. The lack of elongation or biconical structures either in the morphology of the ionized gas or in the continuum tracing the dominant quasar illumination directions is one of the major surprises of our observations. One possible explanation is that the typical opening angles of quasar illumination are large, producing quasi-isotropic illumination pattern. Another possibility is that at the high luminosities explored in this paper the geometry of obscuration is different from that seen in Seyfert galaxies: perhaps there is no organized toroidal structure, but rather obscuration is quasi-spherical and patchy, so that ionizing emission escapes along random beams (but not toward the observer) which we cannot spatially resolve by our data. Both these possibilities are at odds with our \emph{HST} imaging of somewhat less luminous type 2 quasars which clearly shows ionization cones in at least some objects \citep{zaka06}, although a strong luminosity dependence of the geometry of obscuration would reconcile these results. \emph{HST} imaging of the objects in this sample in the blue band sensitive to the scattered light emission can settle this issue. 

An alternative possibility is that shock excitation rather than photo-ionization is responsible for the observed strong line emission. The radiation pressure of the quasar \citep{murr95, prog00} drives a low-density high-temperature wind \citep{fauc12, zubo12} which slams into dense clumps in the host galaxy, resulting in shocked emission. Numerical simulations demonstrate that even if there are anisotropies in the initial driving of the outflow, the outflow isotropizes over the range of scales involved between the scale on which the outflow is accelerated ($\la 10$ pc) and the scale on which it sweeps up the interstellar medium of the galaxy ($\sim 10$ kpc), producing structures qualitatively similar to those seen in our observations \citep{nova11}. Shocks propagating at several hundred km s$^{-1}$ normally do not produce \oiii/H$\beta\sim 10$ \citep{rich11} as seen in our quasars, and thus they are normally disfavored as the primary origin of narrow line emission. Furthermore, the efficiency of converting mechanical energy into line emission is expected to be low at such shock velocities \citep{nesv08}. One possibility is that the shock blast propagates much faster ($>1000$ km s$^{-1}$) than was previously considered, resulting in higher \oiii/H$\beta$ ratios consistent with those observed. Thus, we will continue to explore this possibility in our future work since photo-ionization does not at the moment provide a fully satisfactory explanation of all of the aspects of our data.

\subsection{Physical conditions in the nebulae}
\label{sec:model}

In this Section we discuss a plausible model of a nebula under the assumption that the gas is illuminated and photo-ionized in a wide-angle manner, either because of the large opening angles in the obscuring material or because of the quasi-spherical patchy illumination. Unless noted otherwise, the arguments in this section apply both to a static nebula and to an outflow. Throughout this section we use $r$ to denote the three-dimensional distance from the quasar and $R$ to denote the distance as seen projected onto the plane of the sky.

We aim to determine plausible physical conditions for the line-emitting 
gas that we observe.  Regardless of the mechanism by which the gas 
reached the observed radii, it is reasonable to draw an analogy with 
galaxy-scale winds, in which there is also observed warm line-emitting 
gas at large radii.  In these cases, the gas is multi-phase, 
with different phases of the medium observable 
in different domains of the electromagnetic spectrum. By analogy with the winds 
driven by supernova explosions in star-burst galaxies \citep{heck90, veil05}, the simplest 
model involves two components. One consists of hot, low-density, volume-filling 
potentially X-ray-emitting plasma, whereas the other is warm ($T\sim 10^4$ K), 
higher density component concentrated in clumps, shells or filaments (hereafter 
``clouds''). It is the latter that produces the emission lines that we see in optical 
spectra. If the two components are in rough pressure equilibrium with each other, 
their densities and temperatures are related via 
$n_{\rm hot} T_{\rm hot} \simeq n_{\rm warm} T_{\rm warm}$. 

The \oiii/H$\beta$ ratio shows a remarkably uniform behavior as a function of 
distance from the center across our sample. As Figure~\ref{fig:o3hb} demonstrates, 
for all the radio and radio-quiet objects that we observe (except for 
SDSS~J0319-0019 where we have only a weak detection of H$\beta$), the ratio remains constant 
at \oiii/H$\beta=12.3\pm2.7$ out to distances of $7.0\pm1.8$ kpc, and declines 
as $\log$ \oiii/H$\beta\propto R^{-3.3\pm1.1}$ at larger distances. 
In addition, most objects show a characteristic excess ($\sim$0.2 dex) of 
\oiii/H$\beta$ right at about the break radius. 

\begin{figure*}
\centering
\includegraphics[scale=0.8,clip,trim=0cm 19cm 0cm 0cm]{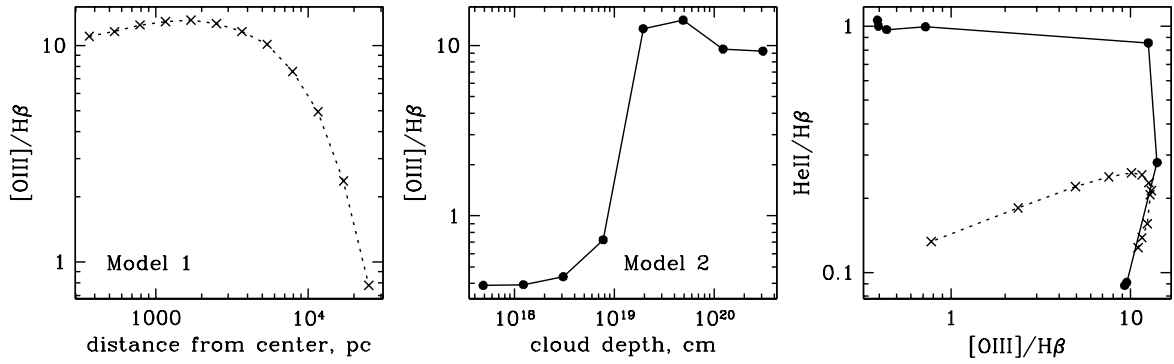}
\caption{Left: In the model where all clouds are ionization-bounded and have the 
same density ($n_{\rm H}=100$ cm$^{-3}$), the \oiii/H$\beta$ ratio shows a behavior 
qualitatively similar to that observed in our sources (plotted here as a function of three-dimensional distance $r$). Center: A model with declining pressure profile uses the fact that the \oiii/H$\beta$ ratio 
changes dramatically as the clouds transition from being ionization-bounded to 
matter-bounded. Here, we show the \oiii/H$\beta$ ratios of clouds of varying thickness 
at a distance of 1 kpc from a $10^{46}$ erg s$^{-1}$ quasar calculated with Cloudy. Right: The two models predict similar \oiii/H$\beta$ behavior as a function of distance, but different \heii/H$\beta$ ratios (Model 1 shown with dashed line, Model 2 shown with solid line). \label{fig:simu}} 
\end{figure*}

We consider two physically distinct models that can qualitatively account for 
the behavior of the line ratios. The first model capitalizes on the fact that 
the \oiii/H$\beta$ ratio is strongly dependent on the ionization parameter 
$U\equiv \Phi/4\pi c r^2 n_{\rm H}(r)$, where $\Phi$ is the photon luminosity of the 
quasar above the ionization energy of hydrogen of 13.6 eV. Following \citealt{vill08}, 
we consider a simple case of clouds of constant density $n_{\rm H}=100$ cm$^{-3}$ 
illuminated by a quasar with bolometric luminosity $L_{\rm bol}=10^{46}$ erg s$^{-1}$. 
In this model, for a typical quasar spectral energy distribution \citep{ferl98} the 
ionization parameter as a function of distance is $\log U=-0.87-\log(r/{\rm kpc})^2$. 
We then use calculations presented in \citealt{vill08} to determine the \oiii/H$\beta$ 
ratios of such clouds (Figure~\ref{fig:simu}, left panel labeled ``Model 1''). As 
long as the clouds remain ``ionization-bounded'' at all distances from the quasar (i.e., the number of O$^{2+}$-ionizing photons is insufficient to ionize the entire cloud) and the value of $U$ 
stays above $\log U\simeq -2.3$, the \oiii/H$\beta$ ratio is constant and close to 10. As $r$ increases and $U$ declines below $\log U \simeq -2.3$, the \oiii/H$\beta$ ratio declines as well, in qualitative
agreement with our observations.

Despite the apparent success of this model in reproducing the \oiii/H$\beta$ ratio, 
the input physics is not entirely satisfactory. In particular, since the clouds' 
thermal balance is maintained by photo-ionization and recombination, their temperatures 
are nearly constant under a wide range of conditions with $T=f\times 10^4$ K where 
the factor $f$ is typically between 1 and 2 \citep{krol99}. Since the model assumes 
a constant density inside the clouds, this implies that the cloud pressure is independent 
of the distance from the quasar as well. If the clouds are in pressure equilibrium with 
the confining low-density medium, this would imply that the pressure of the entire 
quasar nebula is constant as a function of radius. This is hard to reconcile with models, as well as with observations of starburst-driven winds in which pressure declines outwards \citep{heck90}. 

Therefore, we consider another simple model in which the pressure in the nebula 
follows $P(r)\propto r^{-2}$. The clouds are in confined by the pressure of the hot low-density component, and since their temperature 
is almost constant, the density inside them drops as $n_{\rm H}(r)\propto r^{-2}$ as well, 
which implies that the ionization parameter is independent of the distance from the 
quasar. We use the photo-ionization code Cloudy \citep{ferl98} to model emission of 
clouds with a hydrogen density of $n=100$ cm$^{-3}$ at $r=1$ kpc from a 
$L_{\rm bol}=10^{46}$ erg s$^{-1}$ quasar. These parameters correspond to an ionization 
parameter of $\log U=-0.87$.

In Figure~\ref{fig:simu}, center, we show the \oiii/H$\beta$ ratios of such clouds as 
a function of their physical depth. Large clouds are ionization-bounded (i.e., have
a large optical depth) and show the familiar ratio of \oiii/H$\beta\simeq 10$. 
Smaller clouds, however, allow for the photons with energies above 54.9 eV (the energy 
required to ionize O$^{2+}$ into O$^{3+}$) to penetrate through the entire cloud, 
destroying most O$^{2+}$ ions and resulting in much lower \oiii/H$\beta$ ratios. If a cloud is participating in the outflow, it expands (linear size $S \propto r^{2/3}$), 
its density falls ($\propto r^{-2}$) and its optical depth declines ($\propto r^{-4/3}$). 
Therefore, as the clouds propagate out with the quasar wind, they are expected to transition 
from ionization-bounded (partly ionized) to matter-bounded (fully ionized). The same transition occurs in a static nebula as long as clouds have the same mass distribution at all distances. As a result, 
the \oiii/H$\beta$ ratio is expected to fall as a function of distance from the quasar, in 
qualitative agreement with our observations. 

All the arguments presented thus far apply equally to a static nebula or a wind, as long as the pressure profiles are similar. We now consider specifically an outflowing wind. In the matter-bounded regime the 
H$\beta$ luminosity of each cloud $L_{\rm cloud}\propto n^2 S^3 \propto r^{-2}$. If the number 
density of clouds falls off as $r^{-2}$ (as expected in a steady-state constant velocity wind), then the H$\beta$ luminosity density falls off as $r^{-4}$. As a result, the surface brightness of the nebula is expected to fall off as 
$R^{-3}$, which is close to the values seen in our sample (Table~\ref{tab2}, the mean value
of the exponent is $-3.5\pm1.0$). 

The major difference between Models 1 and 2 is in the degree of ionization of clouds at large distances from the quasar. In Model 1, the ionization parameter decreases outward and the clouds have a smaller degree of ionization than they do in the inner parts. In Model 2, the hard radiation can penetrate through the entire cloud, ionizing O$^{2+}$ to higher ionization states. Coincidentally, the ionization energy of O$^{2+}$ (54.9 eV) is almost identical to the ionization energy of He$^{+}$ (54.4 eV). Thus the same photons that destroy O$^{2+}$ and lead to the decrease of the \oiii/H$\beta$ ratio in Model 2 produce He$^{2+}$. This leads to an increase in He$^{2+}$ recombination emission, including one of the transitions in the recombination cascade observed as the optical line \heii\ $\lambda$4686\AA. Thus, Model 2 predicts a much higher ratio of \heii/H$\beta$ in the outer parts in the nebula than does Model 1. The model predictions are illustrated in Figure \ref{fig:simu}, right. The models do not take into account projection effects -- along any given line of sight, regions at different distance $r$ from the quasar contribute to the emission. Nevertheless, it is clear that high \heii/H$\beta$ ratios, especially in combination with low \oiii/H$\beta$, are characteristic of the high-ionization matter-bounded regions and can only be produced in Model 2.

In Figure \ref{fig:he2} we present the measurements of the \heii/H$\beta$ line ratios in the 10 sources where the 
\heii\ $\lambda$4686\AA\ transition is covered by our data. The \heii\ line is weaker than H$\beta$ and is thus cannot be measured out to the same distances from the quasar, although typically we are able to detect \heii\ at distances comparable to or larger than the break radius. There is a general tendency for \heii/H$\beta$ ratio to increase outward, and in 9/10 objects it goes above $\log$ [\heii/H$\beta$]=$-0.5$ -- too high to be compatible with Model 1. The key upper left region of the \heii/H$\beta$ vs \oiii/H$\beta$ diagnostic diagrams is particularly observationally difficult because it is reached only in the outermost parts of the nebulae where \oiii\ is faint; nevertheless, the line ratios in two objects -- SDSS~J0210$-$1001 and SDSS~J0842$+$3625 -- extend into this region. 

We will continue to refine Models 1 and 2 as we investigate other aspects of our data. From the simple model setup presented above, we conclude that the pressure in the nebulae is most likely decreasing outward and that line-emitting clouds transition from ionization-bounded to matter-bounded as the distance from the quasar increases.

\begin{figure*}
\centering
\includegraphics[totalheight=3cm,angle=0,origin=c,scale=0.92]{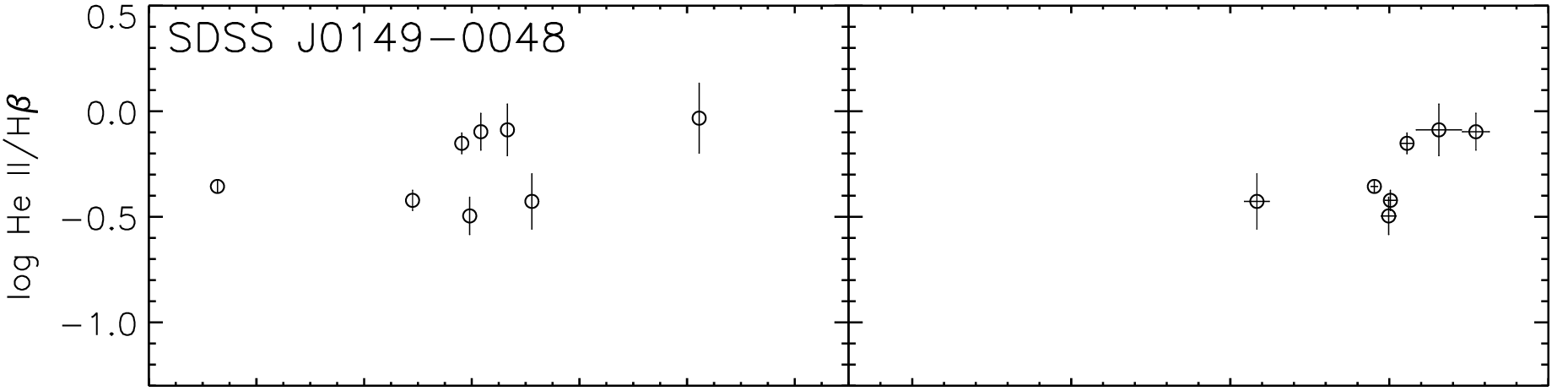}
\includegraphics[totalheight=3cm,angle=0,origin=c,scale=0.92]{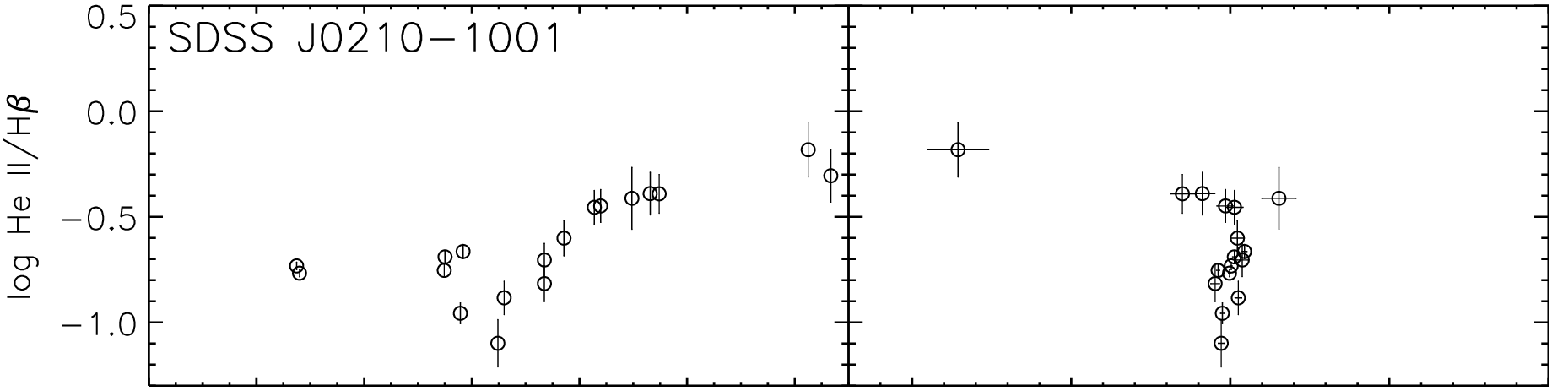}
\includegraphics[totalheight=3cm,angle=0,origin=c,scale=0.92]{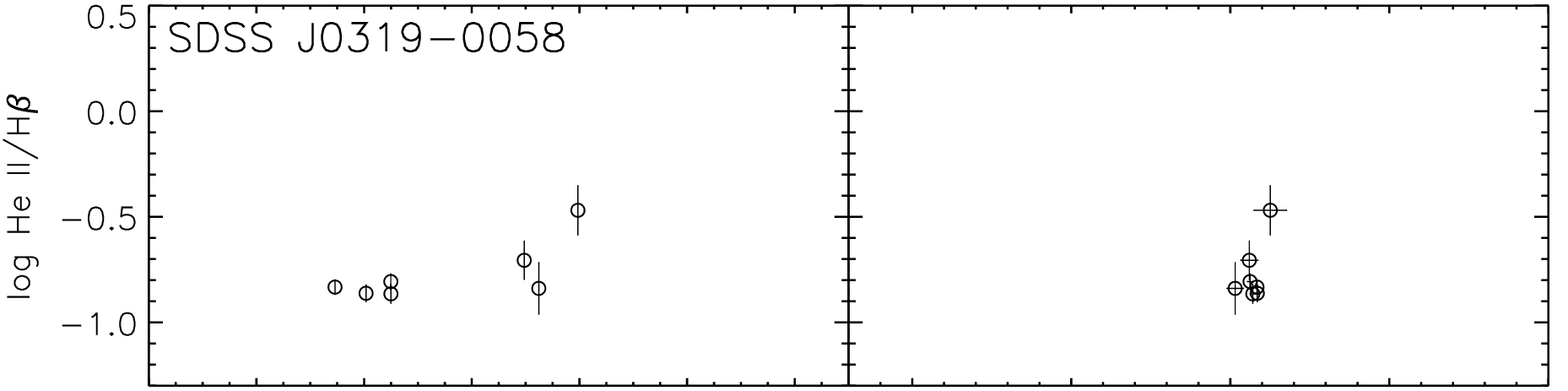}
\includegraphics[totalheight=3cm,angle=0,origin=c,scale=0.92]{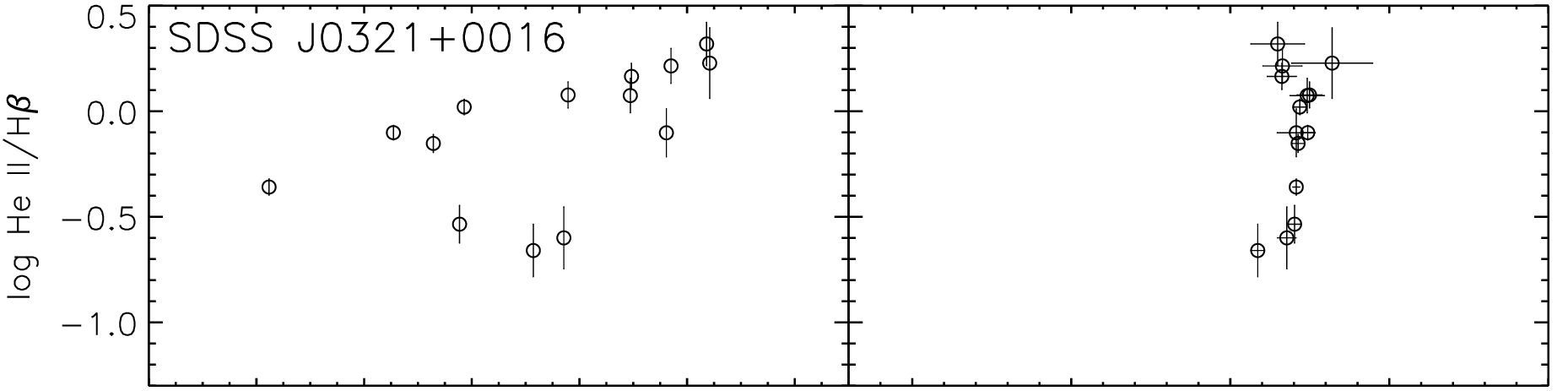}
\includegraphics[totalheight=3cm,angle=0,origin=c,scale=0.92]{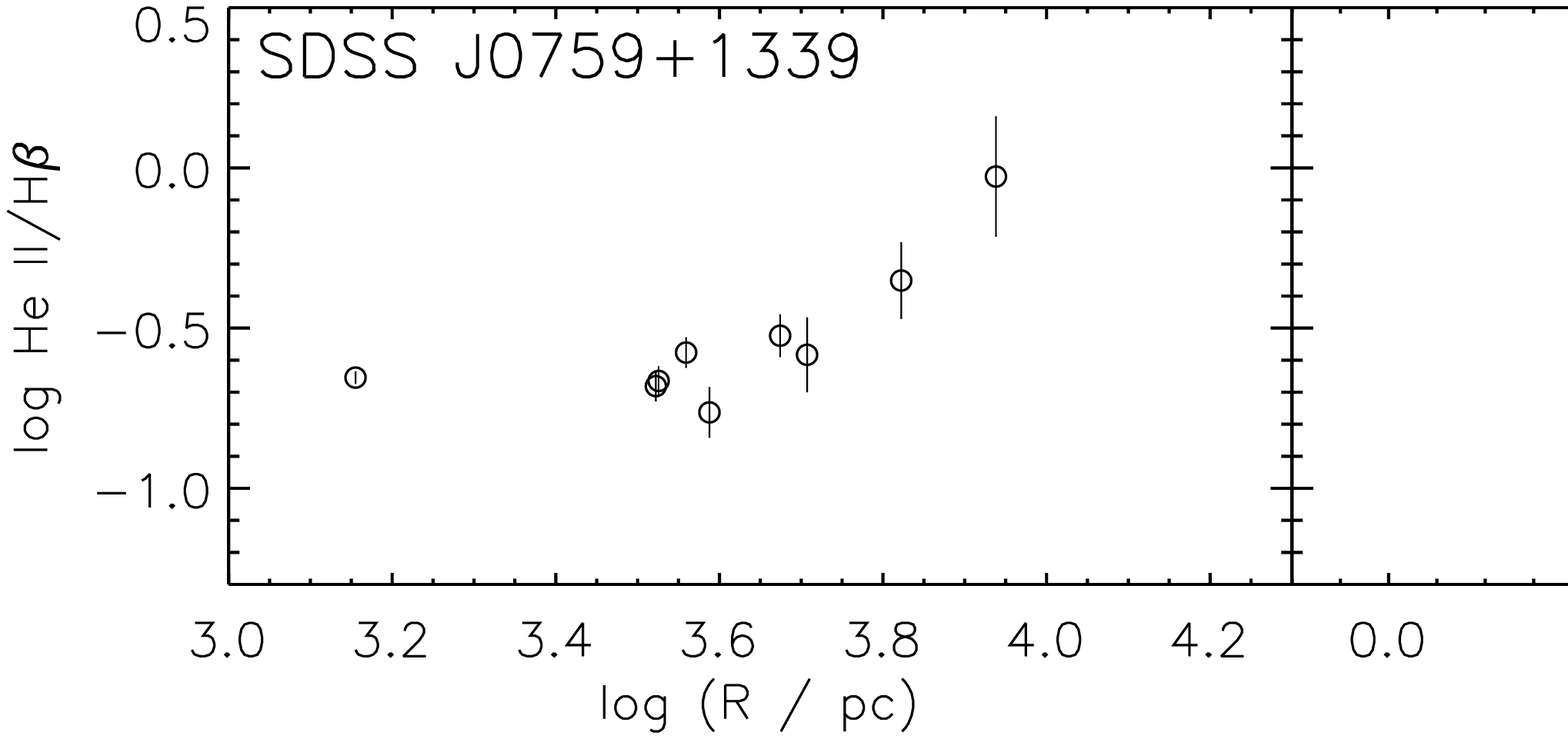}\\
\vspace{1cm}
\caption{He {\sc ii}/H$\beta$ as a function of \oiii\ isophotal radius (left panels)
and \oiii/H$\beta$. Large values of \heii/H$\beta$ ($\log$ \heii/H$\beta\ga-0.5$) strongly suggest presence of matter-bounded clouds, especially in combination with low \oiii/H$\beta$ ($\log$ \oiii/H$\beta\la 0.5$), as seen in SDSS~J0210$-$1001 and SDSS~J0842$+$3625.
}
\label{fig:he2}
\end{figure*}
\clearpage

\begin{figure*}
\centering
\includegraphics[totalheight=3cm,angle=0,origin=c,scale=0.92]{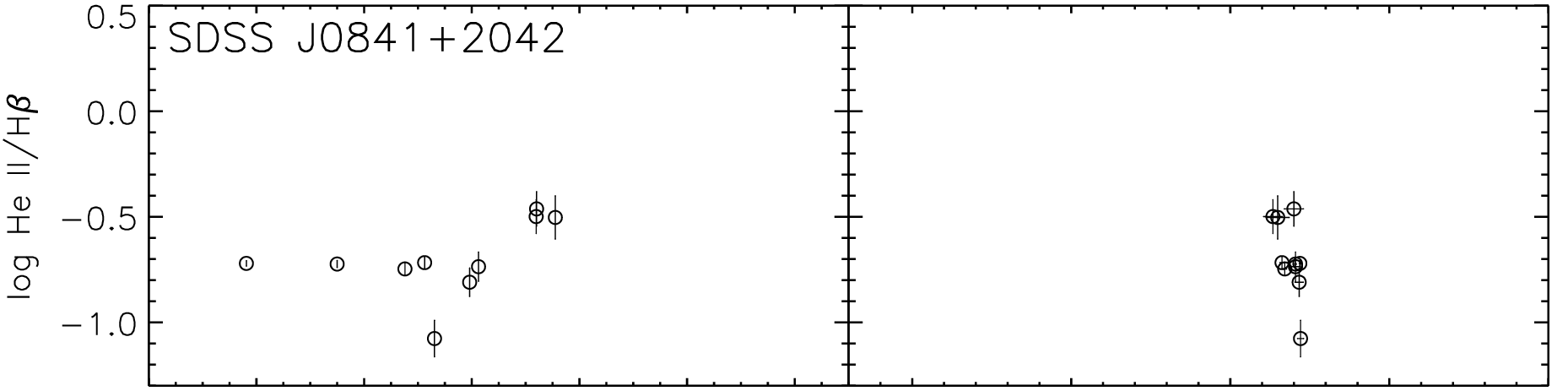}
\includegraphics[totalheight=3cm,angle=0,origin=c,scale=0.92]{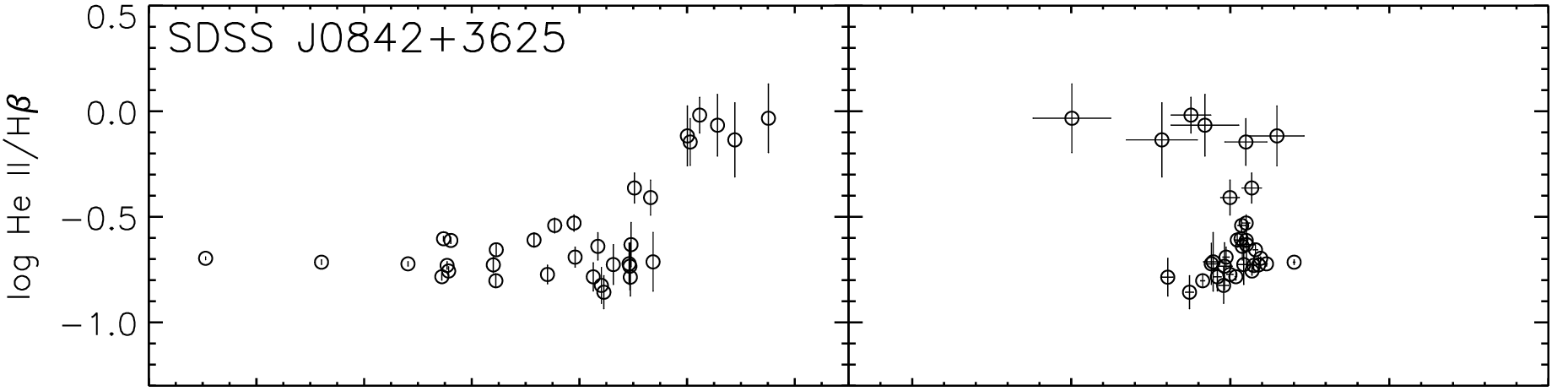}
\includegraphics[totalheight=3cm,angle=0,origin=c,scale=0.92]{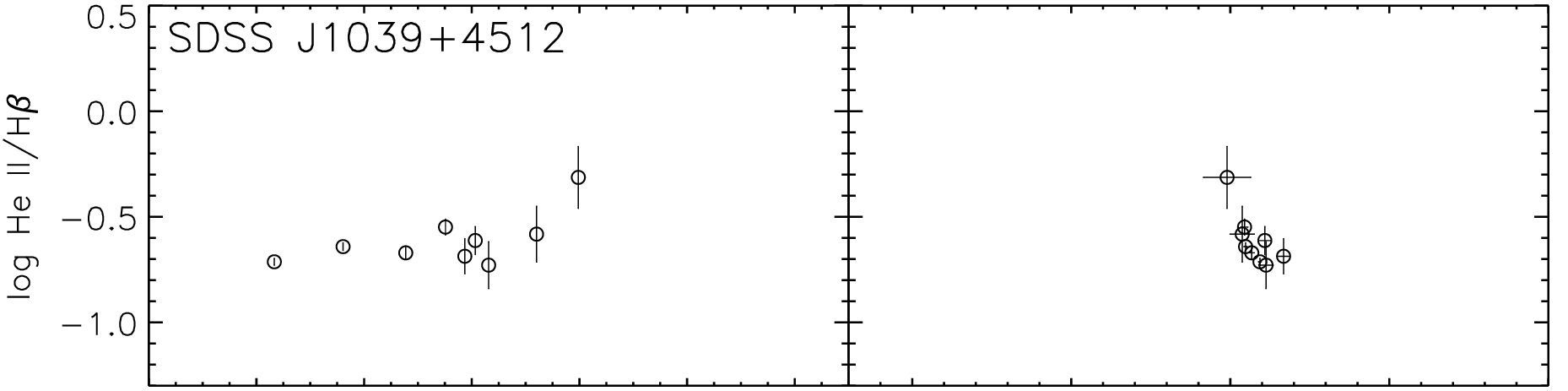}
\includegraphics[totalheight=3cm,angle=0,origin=c,scale=0.92]{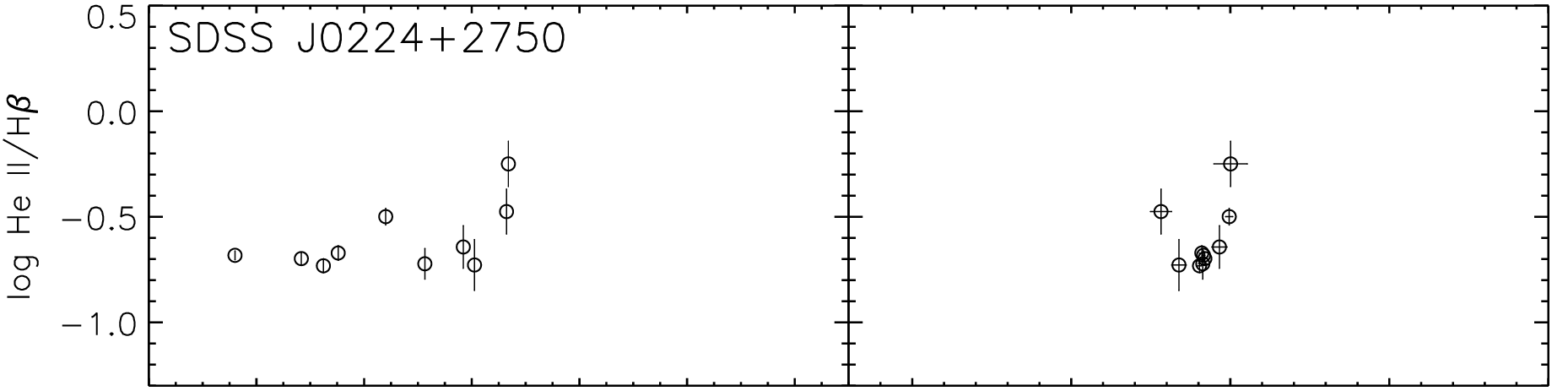}
\includegraphics[totalheight=3cm,angle=0,origin=c,scale=0.92]{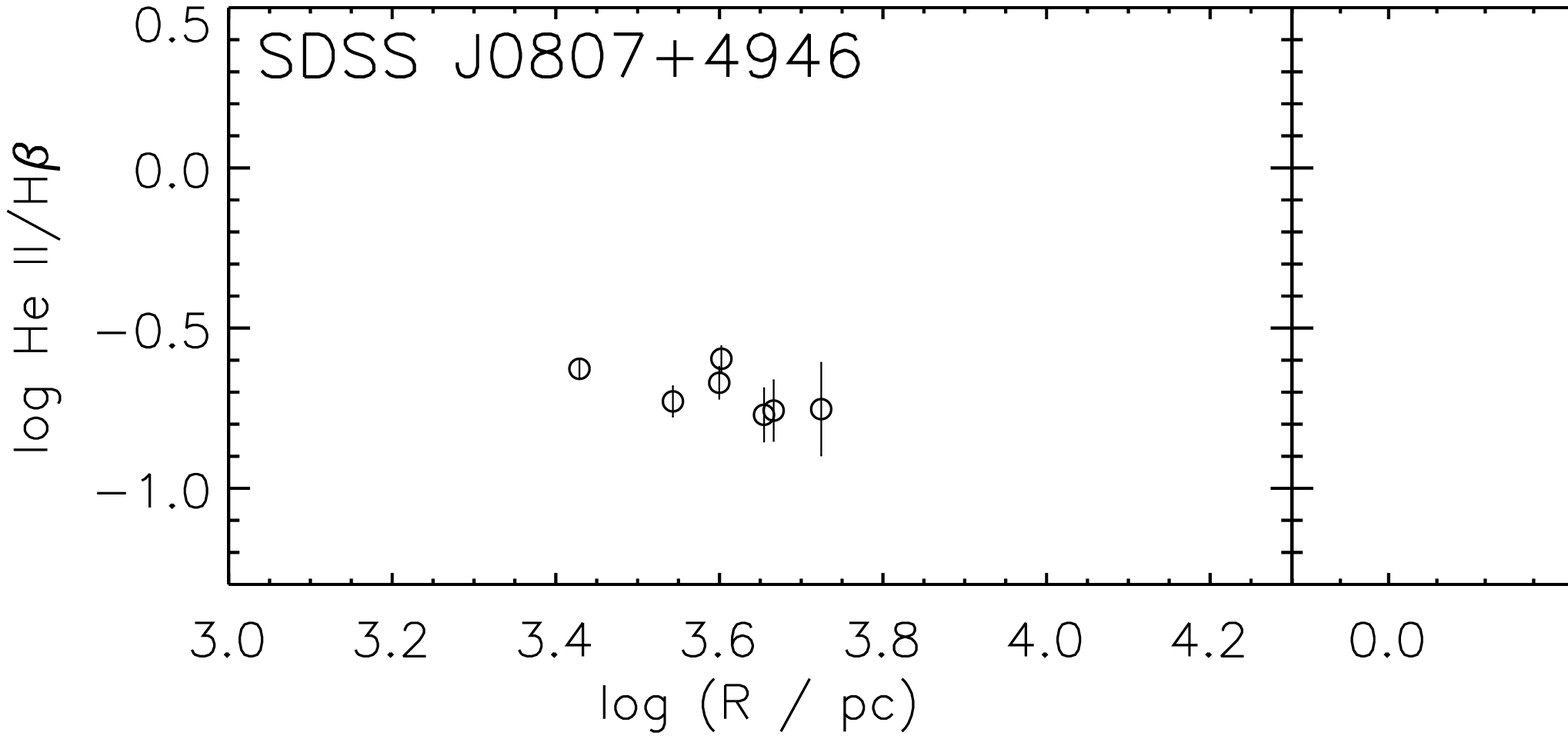}\\
\vspace{1cm}
\caption{Continued.}
\end{figure*}
\clearpage

\section{Conclusions}
\label{sec:conclusions}

In this paper we present optical seeing-limited integral field unit observations of 
ionized gas around $z\simeq 0.5$ obscured luminous quasars. We use these data to 
determine the spatial distribution of the \oiii$\lambda$5007\AA\ and H$\beta$ line emission. 
We spatially resolve the emission line nebulae in every case and find that the \oiii\ 
line emission from gas photo-ionized by the hidden quasar is detected out to $14\pm4$ kpc from the center of the galaxy
(sample mean and standard deviation are given throughout this section). Ionized gas nebulae 
around radio-quiet obscured quasars display regular smooth morphologies, in marked contrast 
with nebulae around radio-loud quasars of similar line luminosities which tend to be 
significantly more elongated and/or lumpy. Surprisingly, no pronounced biconical 
structures expected in a simple quasar illumination model are detected. 

The main measurements we present in this paper are (1) the spatial distribution of the 
emission-line surface brightness, (2) the spatial distribution of the \oiii/H$\beta$ 
ratios, and (3) the relationship between the size of the nebula and the emission line 
luminosity. When fit to a power law, the surface brightness of H$\beta$ declines as 
$R^{-3.5\pm 1.0}$ as a function of the distance from the quasar,
whereas \oiii\ surface brightness declines as $R^{-3.8\pm 0.7}$. The \oiii/H$\beta$ 
ratios show a remarkably uniform behavior across the sample: the ratio remains constant 
at $12.3 \pm 2.7$ and at $R\ga 7\pm2$ kpc declines as roughly a power-law function of the 
distance from the quasar with an index $-3.3 \pm 1.1$. Finally, we supplement our sample 
with 31 objects in the literature, define a redshift- and sensitivity-independent nebula 
size and present a size-luminosity relationship uniformly measured over six orders of 
magnitude in line luminosity. The nebular size slowly increases with luminosity 
$\propto L^{0.250\pm 0.018}$. 

The galaxy-wide narrow-line emission regions around our quasars are large, but the origin of this ionized gas is unclear. In principle, the quasar may be illuminating previously existing gas in the galaxy halo \citep[e.g. from previous star formation,][]{stei10}; or the gas may be debris associated with recent merger activity and currently illuminated by the quasar \citep[see, e.g.][]{fu09, stoc06}. The remarkable roundness of the nebulae and the lack of distinct blobs or clumps is in direct contrast to the morphologies seen in radio-loud objects \citep[c.f.][]{fu11}. It is hard to understand how tidal debris or gas accretion could leave such a smooth and isotropic distribution of gas. One option is that the gas is being expelled from the galaxy, presumably driven by the radiative pressure of the quasar itself \citep{murr95, prog00}. We cannot rule out that previous generations of star formation have left behind the gas now illuminated by the quasar, but we do note that the smooth featureless morphologies of the \oiii\ nebulae in our sample are similar to those seen in wide-angle outflows recently observed by \citet{rupk11}.

We put forward a simple model of a clumpy gaseous nebula, in which 
the lower density hotter component is largely invisible, but the higher-density warm 
component is the one we observe via line emission. In our fiducial model, the gas pressure 
inside the nebula and in the clouds drops off as $r^{-2}$ and the temperature of the 
emission-line clouds is nearly constant at a few $\times 10^4$ K determined by the 
photo-ionization and emission balance. The clouds closer to the quasar are denser and 
are ionization-bounded (i.e., the photo-ionizing photons from the quasar are all absorbed 
on the quasar-facing side of the cloud and cannot penetrate all the way), whereas the 
clouds beyond $r\ga 7$ kpc are lower-density, larger and matter-bounded (i.e., the 
recombination rate cannot keep up with the ionization rate, and these clouds are fully 
ionized). Unlike the model with a constant pressure profile, the model with a declining 
pressure profile qualitatively explains our \oiii/H$\beta$ and \heii/H$\beta$ line ratio 
observations and the surface brightness distributions, and will be tested and refined 
further by our on-going observations and modeling. 

The kinematic analysis of extended line emission around quasars is on-going and will be 
presented in our upcoming papers. We will demonstrate that the nebulae whose morphological 
properties are discussed here are likely due to ionized gas in the outflow from the quasar. 
If these outflows are found to carry significant kinetic energy and/or momentum ($\sim 1\%$ 
of the bolometric output of the quasar), they will be excellent candidates for the long-sought 
feedback phase of quasar evolution. 

\acknowledgments

We thank M. Swinbank and M. Westmoquette for help with data reduction. We acknowledge useful discussions with R. Cen, E. Choi, V. Gaibler, J. Krolik, M.C. Miller, N. Murray, G.S. Novak, J.P. Ostriker, J. Silk, A. Socrates, S. Tremaine, S. Veilleux, and the anonymous referee. N.L.Z. is grateful to the Institute for Advanced Study in Princeton, NJ (where part of this work was performed) for hospitality. N.L.Z. is supported in part by the Alfred P. Sloan fellowship. G.L. and N.L.Z. acknowledge support from the Theodore Dunham, Jr. Grant of the Fund for Astrophysical Research. Support for the work of X.L. was provided by NASA through Einstein Postdoctoral Fellowship grant number PF0--110076 awarded by the {\it Chandra} X-ray Center, which is operated by the Smithsonian Astrophysical Observatory for NASA under contract NAS8--03060.

\clearpage

\begin{deluxetable}{ccrrrccccc}
\tabletypesize{\small}
\rotate
\tablecolumns{10}
\tablewidth{7.6 in}
\tablecaption{Properties of our Gemini quasar sample.\label{tab1}}
\tablehead{
\colhead{Object name} & \colhead{Radio} 	& \colhead{$f_{\rm1.4 GHz}$} & \colhead{$z$}	
& \colhead{PA} & \colhead{$t_{\rm exp}$}	& \colhead{Seeing} & \colhead{$\log L_{\rm [OIII]}$} 
& \colhead{$\delta L/L$} & \colhead{$\eta$} 
\\ 
\colhead{(1)} & \colhead{(2)} & \colhead{(3)} & \colhead{(4)} & \colhead{(5)} & \colhead{(6)} & \colhead{(7)} 
& \colhead{(8)} & \colhead{(9)} & \colhead{(10)} 
}
\startdata

SDSS J014932.53$-$004803.7  & RQ &    1.03 &	0.566 &	323	& 1800$\times$2	& 0.52 & 42.87 & 0.13 & 5.70$\pm$0.14 \\
SDSS J021047.01$-$100152.9  & RQ & $<$1.20 &	0.540 &	280	& 1800$\times$2	& 0.58 & 43.48 & 0.10 & 3.00$\pm$0.02 \\
SDSS J031909.61$-$001916.7  & RQ &    1.95 &	0.635 &	273	& 1800$\times$2	& 0.38 & 42.74 & 0.16 & 3.00$\pm$0.20 \\
SDSS J031950.54$-$005850.6  & RQ & $<$0.97 &	0.626 &	220	& 1800$\times$2	& 0.58 & 42.96 & 0.11 & 4.30$\pm$0.11 \\
SDSS J032144.11$+$001638.2  & RQ &    1.58 &	0.643 &	150	& 1800$\times$2	& 0.54 & 43.10 & 0.11 & 3.02$\pm$0.08 \\
SDSS J075944.64$+$133945.8  & RQ &    1.82 &	0.649 &	164	& 1800$\times$2	& 0.61 & 43.38 & 0.08 & 3.39$\pm$0.02 \\
SDSS J084130.78$+$204220.5  & RQ & $<$0.95 &	0.641 &	250	& 1800$\times$2	& 0.52 & 43.31 & 0.10 & 3.84$\pm$0.02 \\
SDSS J084234.94$+$362503.1  & RQ &    1.64 &	0.561 &	0	& 1800$\times$2	& 0.46 & 43.56 & 0.14 & 3.93$\pm$0.02 \\
SDSS J085829.59$+$441734.7  & RQ & $<$0.90 &	0.454 &	55	& 1800$\times$2	& 0.60 & 43.30 & 0.06 & 4.06$\pm$0.02 \\
SDSS J103927.19$+$451215.4  & RQ &    2.13 &	0.579 &	90	& 1800$\times$2	& 0.58 & 43.29 & 0.13 & 4.00$\pm$0.02 \\
SDSS J104014.43$+$474554.8  & RQ &    4.28 &	0.486 &	10	& 1800$\times$2	& 0.74 & 43.52 & 0.09 & 3.53$\pm$0.01 \\
3C67/J022412.30$+$275011.5  & RL &    3024 & 0.311 & 264	& 1800$\times$2 	& 0.52 & 42.83 & 0.25 & 3.38$\pm$0.06 \\
SDSS J080754.50$+$494627.6  & RL &   169.2 & 0.575 & 152	& 1800$\times$2 	& 0.63 & 43.27 & 0.16 & 3.49$\pm$0.19 \\
SDSS J110140.54$+$400422.9  & RI &   17.38 & 0.457 &	77	& 1800$\times$1	& 0.63 & 43.55 & 0.09 & 4.64$\pm$0.12 \\

\enddata

\begin{tablenotes}{}\scriptsize

\item {\bf Notes.} -- 
(1) object name. 
(2) Radio loudness (RQ: radio quiet; RL: radio loud; RI: radio intermediate).
(3) Radio flux at 1.4 GHz in mJy, taken from the FIRST survey \citep{beck95,whit97} and the NVSS survey
\citep[][for 3C67 only]{cond98}. Upper limits are reported for undetected sources (see text).
(4) Redshift, from \citet{zaka03} and \citet{reye08} for SDSS objects and \citet{erac04} for 3C67. 
(5) Position angle of the field of view (in degrees), measured by the shorter axis from north to east.
(6) Exposure time (in seconds) and number of exposures.
(7) Seeing at the observing site (FWHM, in arcseconds).
(8) Total luminosity of the [O {\sc iii}]$\lambda$5007\AA\ line (erg s$^{-1}$), derived from our data calibrated 
	against SDSS spectra, except for 3C67 whose [O {\sc iii}] flux is taken from \citet{Geld94}.
(9) Relative uncertainty of the \oiii\ luminosity. The uncertainty for 3C67 is taken from \citet{Geld94}.
(10) Absolute value of the best-fit power-law exponent of the outer part of the \oiii\ profile along the major 
axes (i.e., $I_{R,{\rm [OIII]}}\propto R^{-\eta}$, see Figure~\ref{fig:psf}).

\end{tablenotes}

\end{deluxetable}

\clearpage

\begin{deluxetable}{cccrrccrrrrccccc}
\tabletypesize{\small}
\rotate
\tablecolumns{15}
\tablewidth{8.6 in}
\tablecaption{Properties of our Gemini quasar sample (continued).\label{tab2}}
\tablehead{
\colhead{Object} 	& \colhead{$R_{5\sigma}^{\rm cont}$} 		& \colhead{$\epsilon_{5\sigma}^{\rm cont}$}
& \colhead{$R_{5\sigma}$} & \colhead{$\epsilon_{5\sigma}$} & \colhead{$R_{\rm eff}^{\rm cont}$} & \colhead{$R_{\rm eff}$}
& \colhead{$R_{\rm obs}$} & \colhead{$\epsilon_{\rm obs}$} & \colhead{$R_{\rm int}$}	& \colhead{$\epsilon_{\rm int}$}
& \colhead{const} & \colhead{$R_{\rm br}$}	 & \colhead{slope} & \colhead{$\zeta$}
\\
\colhead{(1)} & \colhead{(2)} & \colhead{(3)} & \colhead{(4)} & \colhead{(5)} & \colhead{(6)} & \colhead{(7)} & 
\colhead{(8)} & \colhead{(9)} & \colhead{(10)} & \colhead{(11)} & \colhead{(12)} & \colhead{(13)} & 
\colhead{(14)} & \colhead{(15)}}

\startdata

SDSS J0149$-$0048  & 6.3 & 0.06 &  9.1 & 0.10 & 3.1 & 2.9 & 7.4  & 0.12 &  6.9 & 0.05 & 31.8 & 4.1  & $-$4.79$\pm$0.37 & 3.27$\pm$0.11 \\
SDSS J0210$-$1001  & 4.8 & 0.21 & 17.7 & 0.07 & 3.0 & 4.0 & 17.8 & 0.14 & 14.4 & 0.12 & 10.0 & 7.5  & $-$3.09$\pm$0.22 & 2.08$\pm$0.14 \\
SDSS J0319$-$0019  & 9.9 & 0.13 &  7.6 & 0.25 & 3.0 & 2.7 & 7.9  & 0.22 &  6.7 & 0.23 &  --- & ---  &         ---      &     ---       \\
SDSS J0319$-$0058  & 5.5 & 0.16 & 11.5 & 0.04 & 3.3 & 3.5 & 11.2 & 0.05 &  9.6 & 0.14 & 11.9 & 7.5  & $-$5.32$\pm$1.34 & 4.64$\pm$0.22 \\
SDSS J0321$+$0016  & 8.2 & 0.17 & 18.3 & 0.42 & 4.3 & 5.2 & 19.3 & 0.38 & 16.6 & 0.42 & 16.2 & 11.0 & $-$2.66$\pm$0.35 & 5.02$\pm$0.23 \\
SDSS J0759$+$1339  & 6.0 & 0.22 & 14.1 & 0.02 & 3.0 & 3.5 & 15.4 & 0.07 & 12.9 & 0.08 & 16.1 & 7.5  & $-$2.91$\pm$0.31 & 3.06$\pm$0.08 \\
SDSS J0841$+$2042  & 5.5 & 0.07 & 11.9 & 0.09 & 2.8 & 2.5 & 11.8 & 0.10 & 10.0 & 0.17 & 15.7 & 6.4  & $-$3.53$\pm$0.28 & 3.65$\pm$0.08 \\
SDSS J0842$+$3625  & 6.5 & 0.17 & 15.1 & 0.12 & 3.1 & 3.2 & 14.4 & 0.18 & 13.1 & 0.19 & 10.3 & 9.0  & $-$4.14$\pm$0.56 & 4.82$\pm$0.19 \\
SDSS J0858$+$4417  & 6.9 & 0.12 & 11.7 & 0.08 & 3.0 & 3.1 & 11.5 & 0.12 & 10.6 & 0.10 & 11.9 & 5.6  & $-$3.88$\pm$1.16 & 3.24$\pm$0.16 \\
SDSS J1039$+$4512  & 4.9 & 0.19 & 12.2 & 0.24 & 2.5 & 2.6 & 11.4 & 0.22 &  9.3 & 0.12 & 12.2 & 5.8  & $-$3.90$\pm$0.65 & 4.54$\pm$0.14 \\
SDSS J1040$+$4745  & 6.9 & 0.09 & 14.5 & 0.04 & 3.6 & 3.4 & 14.5 & 0.08 & 13.0 & 0.18 & 8.7  & 7.6  & $-$2.45$\pm$0.75 & 2.88$\pm$0.20 \\
3C67/J0224$+$2750  & 5.2 & 0.07 & 12.2 & 0.42 & 2.0 & 2.3 & 19.4 & 0.42 & 10.1 & 0.26 & 4.6  & 8.3  & $-$2.38$\pm$0.90 & 2.99$\pm$0.18 \\
SDSS J0807$+$4946  & 5.8 & 0.22 & 19.4 & 0.44 & 2.7 & 3.3 & 19.2 & 0.47 & 14.6 & 0.45 & 13.8 & 6.3  & $-$1.45$\pm$0.45 & 3.60$\pm$0.04 \\
SDSS J1101$+$4004  & 4.7 & 0.05 & 18.9 & 0.26 & 3.4 & 5.2 & 19.3 & 0.31 & 18.7 & 0.31 & 12.3 & 8.2  & $-$2.90$\pm$0.36 & 2.16$\pm$0.03 \\

\enddata

\begin{tablenotes}{}\scriptsize

\item {\bf Notes.} -- 
(1) object name. 
(2--3) Semi-major axis (in kpc) and ellipticity ($\equiv1-b/a$) of the best-fit ellipse which encloses pixels 
with $S/N\geqslant5$ in the continuum map.
(4--5) Semi-major axis (in kpc) and ellipticity of the best-fit ellipse which encloses pixels 
with $S/N\geqslant5$ in the \oiii$\lambda$5007\AA\ line map.
(6--7) Half-light isophotal radius (effective radius) of the continuum and \oiii$\lambda$5007\AA\ line emission 
(semi-major axis, in kpc).
(8--9) Isophotal radius (semi-major axis, in kpc) and ellipticity at the observed limiting surface brightness of 
$10^{-16}$ erg s$^{-1}$ cm$^{-2}$ arcsec$^{-2}$.
(10--11) Isophotal radius (semi-major axis, in kpc) and ellipticity at the intrinsic limiting surface brightness 
(corrected for cosmological dimming) of $10^{-15}$ erg s$^{-1}$ cm$^{-2}$ arcsec$^{-2}$.
(12) Mean value of the plateau part of the \oiii/H$\beta$ radial profile.
(13) Break radius (semi-major axis) of the \oiii/H$\beta$ radial profile.
(14) Best-fit power-law slope of the \oiii/H$\beta$ radial profile at $R>R_{\rm br}$.
(15) Absolute value of the best-fit power-law exponent of the H$\beta$ radial profile at $R>R_{\rm br}$ (i.e.,  
$I_{R,{\rm H\beta}} \propto R^{-\zeta}$).

\end{tablenotes}

\end{deluxetable}

\begin{deluxetable}{cccccccccc}
\tabletypesize{\footnotesize}
\tablecolumns{10}
\tablewidth{5.82 in}
\tablecaption{Properties of the radio galaxies from \citet{nesv08}.\label{tab3}}
\tablehead{
\colhead{Object name}	& \colhead{Type} & \colhead{$f_{\rm 1.4GHz}$} & \colhead{$z$}	 & \colhead{Seeing}	
& \colhead{$\log L_{\rm [O~{\scriptscriptstyle III}]}$} 	& \colhead{$R_{\rm eff}$}	& \colhead{$\epsilon_{\rm eff}$}	
& \colhead{$R_{\rm int}$}	& \colhead{$\epsilon_{\rm int}$
} \\
\colhead{} & \colhead{} & \colhead{(mJy)} & \colhead{} & \colhead{(\arcsec)} & \colhead{(erg s$^{-1}$)} 
& \colhead{(kpc)} & \colhead{} & \colhead{(kpc)} & \colhead{}
}
\startdata

MRC 0316--257   & RL & 447 & 3.127 & 0.7$\times$0.4 & 44.29 & 4.3 & 0.37 & 21.27 & 0.59  \\
MRC 0406--244   & RL & 626 & 2.425 & 0.7$\times$0.6 & 44.50 & 4.6 & 0.43 & 28.12 & 0.48  \\
TXS 0828+193		& RL & 118 & 2.577 & 0.7$\times$0.5 & 44.51 & 6.6 & 0.56 & 26.66 & 0.61	 \\

\enddata

\begin{tablenotes}{}\scriptsize

\item {\bf Notes.} -- All notations are identical to Table~\ref{tab1} and \ref{tab2}. Redshift (determined from the 
[O {\sc iii}]$\lambda$5007\AA\ line), seeing and [O {\sc iii}] luminosity data are taken from \citet{nesv08}.
Radio fluxes are from the NVSS survey \citep{cond98}.

\end{tablenotes}

\end{deluxetable}

\clearpage

\bibliographystyle{apj}
\bibliography{master}

\begin{thebibliography}{115}
\expandafter\ifx\csname natexlab\endcsname\relax\def\natexlab#1{#1}\fi

\bibitem[{{Abazajian} {et~al.}(2009){Abazajian}, {Adelman-McCarthy},
  {Ag{\"u}eros}, {Allam}, {Allende Prieto}, {An}, {Anderson}, {Anderson},
  {Annis}, {Bahcall}, \& et~al.}]{abaz09}
{Abazajian}, K.~N., {Adelman-McCarthy}, J.~K., {Ag{\"u}eros}, M.~A., {et~al.}
  2009, \apjs, 182, 543

\bibitem[{{Allington-Smith} {et~al.}(2002){Allington-Smith}, {Murray},
  {Content}, {Dodsworth}, {Davies}, {Miller}, {Jorgensen}, {Hook}, {Crampton},
  \& {Murowinski}}]{alli02}
{Allington-Smith}, J., {Murray}, G., {Content}, R., {et~al.} 2002, \pasp, 114,
  892

\bibitem[{{Antonucci}(1993)}]{anto93}
{Antonucci}, R. 1993, \araa, 31, 473

\bibitem[{{Antonucci} \& {Miller}(1985)}]{anto85}
{Antonucci}, R.~R.~J., \& {Miller}, J.~S. 1985, \apj, 297, 621

\bibitem[{{Arav} {et~al.}(2008){Arav}, {Moe}, {Costantini}, {Korista}, {Benn},
  \& {Ellison}}]{arav08}
{Arav}, N., {Moe}, M., {Costantini}, E., {et~al.} 2008, \apj, 681, 954

\bibitem[{{Baldwin} {et~al.}(1981){Baldwin}, {Phillips}, \&
  {Terlevich}}]{bald81}
{Baldwin}, J.~A., {Phillips}, M.~M., \& {Terlevich}, R. 1981, \pasp, 93, 5

\bibitem[{{Baskin} \& {Laor}(2005)}]{bask05o3}
{Baskin}, A., \& {Laor}, A. 2005, \mnras, 358, 1043

\bibitem[{{Baum} {et~al.}(1990){Baum}, {Heckman}, \& {van Breugel}}]{baum90}
{Baum}, S.~A., {Heckman}, T., \& {van Breugel}, W. 1990, \apjs, 74, 389

\bibitem[{{Baum} {et~al.}(1992){Baum}, {Heckman}, \& {van Breugel}}]{baum92}
{Baum}, S.~A., {Heckman}, T.~M., \& {van Breugel}, W. 1992, \apj, 389, 208

\bibitem[{{Becker} {et~al.}(1991){Becker}, {White}, \& {Edwards}}]{beck91}
{Becker}, R.~H., {White}, R.~L., \& {Edwards}, A.~L. 1991, \apjs, 75, 1

\bibitem[{{Becker} {et~al.}(1995){Becker}, {White}, \& {Helfand}}]{beck95}
{Becker}, R.~H., {White}, R.~L., \& {Helfand}, D.~J. 1995, \apj, 450, 559

\bibitem[{{Bennert} {et~al.}(2002){Bennert}, {Falcke}, {Schulz}, {Wilson}, \&
  {Wills}}]{benn02}
{Bennert}, N., {Falcke}, H., {Schulz}, H., {Wilson}, A.~S., \& {Wills}, B.~J.
  2002, \apjl, 574, L105

\bibitem[{{Bennert} {et~al.}(2006){Bennert}, {Jungwiert}, {Komossa}, {Haas}, \&
  {Chini}}]{benn06}
{Bennert}, N., {Jungwiert}, B., {Komossa}, S., {Haas}, M., \& {Chini}, R. 2006,
  \aap, 456, 953

\bibitem[{{Boroson} {et~al.}(1985){Boroson}, {Persson}, \& {Oke}}]{boro85}
{Boroson}, T.~A., {Persson}, S.~E., \& {Oke}, J.~B. 1985, \apj, 293, 120

\bibitem[{{Cavagnolo} {et~al.}(2010){Cavagnolo}, {McNamara}, {Nulsen},
  {Carilli}, {Jones}, \& {B{\^i}rzan}}]{cava10}
{Cavagnolo}, K.~W., {McNamara}, B.~R., {Nulsen}, P.~E.~J., {et~al.} 2010, \apj,
  720, 1066

\bibitem[{{Cecil} {et~al.}(2001){Cecil}, {Bland-Hawthorn}, {Veilleux}, \&
  {Filippenko}}]{ceci01}
{Cecil}, G., {Bland-Hawthorn}, J., {Veilleux}, S., \& {Filippenko}, A.~V. 2001,
  \apj, 555, 338

\bibitem[{{Collinge} {et~al.}(2005){Collinge}, {Strauss}, {Hall}, {Ivezi{\'c}},
  {Munn}, {Schlegel}, {Zakamska}, {Anderson}, {Harris}, {Richards},
  {Schneider}, {Voges}, {York}, {Margon}, \& {Brinkmann}}]{coll05}
{Collinge}, M.~J., {Strauss}, M.~A., {Hall}, P.~B., {et~al.} 2005, \aj, 129,
  2542

\bibitem[{{Condon} {et~al.}(1998){Condon}, {Cotton}, {Greisen}, {Yin},
  {Perley}, {Taylor}, \& {Broderick}}]{cond98}
{Condon}, J.~J., {Cotton}, W.~D., {Greisen}, E.~W., {et~al.} 1998, \aj, 115,
  1693

\bibitem[{{Crenshaw} {et~al.}(2003){Crenshaw}, {Kraemer}, \&
  {George}}]{cren03b}
{Crenshaw}, D.~M., {Kraemer}, S.~B., \& {George}, I.~M. 2003, \araa, 41, 117

\bibitem[{{Croton} {et~al.}(2006)}]{crot06}
{Croton}, D.~J., {et~al.} 2006, \mnras, 365, 11

\bibitem[{{Donahue} {et~al.}(2006){Donahue}, {Horner}, {Cavagnolo}, \&
  {Voit}}]{Dona06}
{Donahue}, M., {Horner}, D.~J., {Cavagnolo}, K.~W., \& {Voit}, G.~M. 2006,
  \apj, 643, 730

\bibitem[{{Dunn} {et~al.}(2010){Dunn}, {Bautista}, {Arav}, {Moe}, {Korista},
  {Costantini}, {Benn}, {Ellison}, \& {Edmonds}}]{dunn10}
{Dunn}, J.~P., {Bautista}, M., {Arav}, N., {et~al.} 2010, \apj, 709, 611

\bibitem[{{Eracleous} \& {Halpern}(2004)}]{erac04}
{Eracleous}, M., \& {Halpern}, J.~P. 2004, \apjs, 150, 181

\bibitem[{{Fanaroff} \& {Riley}(1974)}]{fana74}
{Fanaroff}, B.~L., \& {Riley}, J.~M. 1974, \mnras, 167, 31P

\bibitem[{{Faucher-Gigu{\`e}re} \& {Quataert}(2012)}]{fauc12}
{Faucher-Gigu{\`e}re}, C.-A., \& {Quataert}, E. 2012, \mnras, 425, 605

\bibitem[{{Ferland} {et~al.}(1998){Ferland}, {Korista}, {Verner}, {Ferguson},
  {Kingdon}, \& {Verner}}]{ferl98}
{Ferland}, G.~J., {Korista}, K.~T., {Verner}, D.~A., {et~al.} 1998, \pasp, 110,
  761

\bibitem[{{Ferrarese} \& {Merritt}(2000)}]{ferr00}
{Ferrarese}, L., \& {Merritt}, D. 2000, \apjl, 539, L9

\bibitem[{{Fraquelli} {et~al.}(2003){Fraquelli}, {Storchi-Bergmann}, \&
  {Levenson}}]{fraq03}
{Fraquelli}, H.~A., {Storchi-Bergmann}, T., \& {Levenson}, N.~A. 2003, \mnras,
  341, 449

\bibitem[{{Fu} {et~al.}(2011){Fu}, {Myers}, {Djorgovski}, \& {Yan}}]{fu11}
{Fu}, H., {Myers}, A.~D., {Djorgovski}, S.~G., \& {Yan}, L. 2011, \apj, 733,
  103

\bibitem[{{Fu} \& {Stockton}(2009)}]{fu09}
{Fu}, H., \& {Stockton}, A. 2009, \apj, 690, 953

\bibitem[{{Gallagher} {et~al.}(2007){Gallagher}, {Hines}, {Blaylock},
  {Priddey}, {Brandt}, \& {Egami}}]{gall07}
{Gallagher}, S.~C., {Hines}, D.~C., {Blaylock}, M., {et~al.} 2007, \apj, 665,
  157

\bibitem[{{Ganguly} \& {Brotherton}(2008)}]{gang08}
{Ganguly}, R., \& {Brotherton}, M.~S. 2008, \apj, 672, 102

\bibitem[{{Gebhardt} {et~al.}(2000)}]{gebh00}
{Gebhardt}, K., {et~al.} 2000, \apjl, 539, L13

\bibitem[{{Gelderman} \& {Whittle}(1994)}]{Geld94}
{Gelderman}, R., \& {Whittle}, M. 1994, \apjs, 91, 491

\bibitem[{{Goto} {et~al.}(2003){Goto}, {Nichol}, {Okamura}, {Sekiguchi},
  {Miller}, {Bernardi}, {Hopkins}, {Tremonti}, {Connolly}, {Castander},
  {Brinkmann}, {Fukugita}, {Harvanek}, {Ivezic}, {Kleinman}, {Krzesinski},
  {Long}, {Loveday}, {Neilsen}, {Newman}, {Nitta}, {Snedden}, \&
  {Subbarao}}]{goto03}
{Goto}, T., {Nichol}, R.~C., {Okamura}, S., {et~al.} 2003, \pasj, 55, 771

\bibitem[{{Greene} {et~al.}(2011){Greene}, {Zakamska}, {Ho}, \&
  {Barth}}]{gree11}
{Greene}, J.~E., {Zakamska}, N.~L., {Ho}, L.~C., \& {Barth}, A.~J. 2011, \apj,
  732, 9

\bibitem[{{Greene} {et~al.}(2012){Greene}, {Zakamska}, \& {Smith}}]{gree12}
{Greene}, J.~E., {Zakamska}, N.~L., \& {Smith}, P.~S. 2012, \apj, 746, 86

\bibitem[{{G{\"u}ltekin} {et~al.}(2009)}]{gult09}
{G{\"u}ltekin}, K., {et~al.} 2009, \apj, 698, 198

\bibitem[{{Hao} {et~al.}(2005){Hao}, {Strauss}, {Fan}, {Tremonti}, {Schlegel},
  {Heckman}, {Kauffmann}, {Blanton}, {Gunn}, {Hall}, {Ivezi{\'c}}, {Knapp},
  {Krolik}, {Lupton}, {Richards}, {Schneider}, {Strateva}, {Zakamska},
  {Brinkmann}, \& {Szokoly}}]{hao05}
{Hao}, L., {Strauss}, M.~A., {Fan}, X., {et~al.} 2005, \aj, 129, 1795

\bibitem[{{H{\"a}ring} \& {Rix}(2004)}]{hari04}
{H{\"a}ring}, N., \& {Rix}, H.-W. 2004, \apjl, 604, L89

\bibitem[{{Heckman} {et~al.}(1990){Heckman}, {Armus}, \& {Miley}}]{heck90}
{Heckman}, T.~M., {Armus}, L., \& {Miley}, G.~K. 1990, \apjs, 74, 833

\bibitem[{{Ho}(2005)}]{ho05}
{Ho}, L.~C. 2005, preprint (astroph/0511157)

\bibitem[{{Hopkins} {et~al.}(2006){Hopkins}, {Hernquist}, {Cox}, {Di Matteo},
  {Robertson}, \& {Springel}}]{hopk06}
{Hopkins}, P.~F., {Hernquist}, L., {Cox}, T.~J., {et~al.} 2006, \apjs, 163, 1

\bibitem[{{Humphrey} {et~al.}(2010){Humphrey}, {Villar-Mart{\'{\i}}n},
  {S{\'a}nchez}, {Mart{\'{\i}}nez-Sansigre}, {Gonz{\'a}lez Delgado},
  {P{\'e}rez}, {Tadhunter}, \& {P{\'e}rez-Torres}}]{hump10}
{Humphrey}, A., {Villar-Mart{\'{\i}}n}, M., {S{\'a}nchez}, S.~F., {et~al.}
  2010, \mnras, L113

\bibitem[{{Husemann} {et~al.}(2008){Husemann}, {Wisotzki}, {S{\'a}nchez}, \&
  {Jahnke}}]{huse08}
{Husemann}, B., {Wisotzki}, L., {S{\'a}nchez}, S.~F., \& {Jahnke}, K. 2008,
  \aap, 488, 145

\bibitem[{{Husemann} {et~al.}(2012){Husemann}, {Wisotzki}, {S{\'a}nchez}, \&
  {Jahnke}}]{huse12}
---. 2012, ArXiv e-prints

\bibitem[{{Jia} {et~al.}(2012){Jia}, {Ptak}, {Heckman}, \& {Zakamska}}]{jia12}
{Jia}, J., {Ptak}, A., {Heckman}, T., \& {Zakamska}, N. 2012, ArXiv e-prints

\bibitem[{{Jiang} {et~al.}(2007){Jiang}, {Fan}, {Vestergaard}, {Kurk},
  {Walter}, {Kelly}, \& {Strauss}}]{jian07}
{Jiang}, L., {Fan}, X., {Vestergaard}, M., {et~al.} 2007, \aj, 134, 1150

\bibitem[{{Kelly}(2007)}]{kell07}
{Kelly}, B.~C. 2007, \apj, 665, 1489

\bibitem[{{Krolik}(1999)}]{krol99}
{Krolik}, J.~H. 1999, {Active galactic nuclei : from the central black hole to
  the galactic environment} (Princeton, NJ: Princeton University Press)

\bibitem[{{Lacy} {et~al.}(2007){Lacy}, {Sajina}, {Petric}, {Seymour},
  {Canalizo}, {Ridgway}, {Armus}, \& {Storrie-Lombardi}}]{lacy07}
{Lacy}, M., {Sajina}, A., {Petric}, A.~O., {et~al.} 2007, \apjl, 669, L61

\bibitem[{{Lal} \& {Ho}(2010)}]{lal10}
{Lal}, D.~V., \& {Ho}, L.~C. 2010, \aj, 139, 1089

\bibitem[{{Liu} {et~al.}(2009){Liu}, {Zakamska}, {Greene}, {Strauss}, {Krolik},
  \& {Heckman}}]{liu09}
{Liu}, X., {Zakamska}, N.~L., {Greene}, J.~E., {et~al.} 2009, \apj, 702, 1098

\bibitem[{{Ludwig} {et~al.}(2009){Ludwig}, {Wills}, {Greene}, \&
  {Robinson}}]{ludw09}
{Ludwig}, R.~R., {Wills}, B., {Greene}, J.~E., \& {Robinson}, E.~L. 2009, \apj,
  706, 995

\bibitem[{{Magorrian} {et~al.}(1998){Magorrian}, {Tremaine}, {Richstone},
  {Bender}, {Bower}, {Dressler}, {Faber}, {Gebhardt}, {Green}, {Grillmair},
  {Kormendy}, \& {Lauer}}]{mago98}
{Magorrian}, J., {Tremaine}, S., {Richstone}, D., {et~al.} 1998, \aj, 115, 2285

\bibitem[{{Marconi} \& {Hunt}(2003)}]{marc03}
{Marconi}, A., \& {Hunt}, L.~K. 2003, \apjl, 589, L21

\bibitem[{{McCarthy}(1993)}]{mcca93}
{McCarthy}, P.~J. 1993, \araa, 31, 639

\bibitem[{{McConnell} {et~al.}(2011){McConnell}, {Ma}, {Gebhardt}, {Wright},
  {Murphy}, {Lauer}, {Graham}, \& {Richstone}}]{mcco11}
{McConnell}, N.~J., {Ma}, C.-P., {Gebhardt}, K., {et~al.} 2011, \nat, 480, 215

\bibitem[{{McNamara} \& {Nulsen}(2007)}]{mcna07}
{McNamara}, B.~R., \& {Nulsen}, P.~E.~J. 2007, \araa, 45, 117

\bibitem[{{Moe} {et~al.}(2009){Moe}, {Arav}, {Bautista}, \& {Korista}}]{moe09}
{Moe}, M., {Arav}, N., {Bautista}, M.~A., \& {Korista}, K.~T. 2009, \apj, 706,
  525

\bibitem[{{Morganti} {et~al.}(2005){Morganti}, {Tadhunter}, \&
  {Oosterloo}}]{morg05}
{Morganti}, R., {Tadhunter}, C.~N., \& {Oosterloo}, T.~A. 2005, \aap, 444, L9

\bibitem[{{Murray} {et~al.}(1995){Murray}, {Chiang}, {Grossman}, \&
  {Voit}}]{murr95}
{Murray}, N., {Chiang}, J., {Grossman}, S.~A., \& {Voit}, G.~M. 1995, \apj,
  451, 498

\bibitem[{{Nesvadba} {et~al.}(2008){Nesvadba}, {Lehnert}, {De Breuck},
  {Gilbert}, \& {van Breugel}}]{nesv08}
{Nesvadba}, N.~P.~H., {Lehnert}, M.~D., {De Breuck}, C., {Gilbert}, A.~M., \&
  {van Breugel}, W. 2008, \aap, 491, 407

\bibitem[{{Nesvadba} {et~al.}(2006){Nesvadba}, {Lehnert}, {Eisenhauer},
  {Gilbert}, {Tecza}, \& {Abuter}}]{nesv06}
{Nesvadba}, N.~P.~H., {Lehnert}, M.~D., {Eisenhauer}, F., {et~al.} 2006, \apj,
  650, 693

\bibitem[{{Novak} {et~al.}(2011){Novak}, {Ostriker}, \& {Ciotti}}]{nova11}
{Novak}, G.~S., {Ostriker}, J.~P., \& {Ciotti}, L. 2011, \apj, 737, 26

\bibitem[{{O'Dea} {et~al.}(2002){O'Dea}, {de Vries}, {Koekemoer}, {Baum},
  {Morganti}, {Fanti}, {Capetti}, {Tadhunter}, {Barthel}, {Axon}, \&
  {Gelderman}}]{odea02}
{O'Dea}, C.~P., {de Vries}, W.~H., {Koekemoer}, A.~M., {et~al.} 2002, \aj, 123,
  2333

\bibitem[{{Oosterloo} {et~al.}(2000){Oosterloo}, {Morganti}, {Tzioumis},
  {Reynolds}, {King}, {McCulloch}, \& {Tsvetanov}}]{oost00}
{Oosterloo}, T.~A., {Morganti}, R., {Tzioumis}, A., {et~al.} 2000, \aj, 119,
  2085

\bibitem[{{Osterbrock} \& {Ferland}(2006)}]{oste06}
{Osterbrock}, D.~E., \& {Ferland}, G.~J. 2006, {Astrophysics of gaseous nebulae
  and active galactic nuclei} (Sausalito, CA: University Science Books)

\bibitem[{{Privon} {et~al.}(2008){Privon}, {O'Dea}, {Baum}, {Axon}, {Kharb},
  {Buchanan}, {Sparks}, \& {Chiaberge}}]{priv08}
{Privon}, G.~C., {O'Dea}, C.~P., {Baum}, S.~A., {et~al.} 2008, \apjs, 175, 423

\bibitem[{{Proga} {et~al.}(2000){Proga}, {Stone}, \& {Kallman}}]{prog00}
{Proga}, D., {Stone}, J.~M., \& {Kallman}, T.~R. 2000, \apj, 543, 686

\bibitem[{{Ptak} {et~al.}(2006){Ptak}, {Zakamska}, {Strauss}, {Krolik},
  {Heckman}, {Schneider}, \& {Brinkmann}}]{ptak06}
{Ptak}, A., {Zakamska}, N.~L., {Strauss}, M.~A., {et~al.} 2006, \apj, 637, 147

\bibitem[{{Randall} {et~al.}(2011){Randall}, {Forman}, {Giacintucci}, {Nulsen},
  {Sun}, {Jones}, {Churazov}, {David}, {Kraft}, {Donahue}, {Blanton},
  {Simionescu}, \& {Werner}}]{rand11}
{Randall}, S.~W., {Forman}, W.~R., {Giacintucci}, S., {et~al.} 2011, \apj, 726,
  86

\bibitem[{{Reichard} {et~al.}(2003)}]{reic03}
{Reichard}, T.~A., {et~al.} 2003, \aj, 125, 1711

\bibitem[{{Reyes} {et~al.}(2008){Reyes}, {Zakamska}, {Strauss}, {Green},
  {Krolik}, {Shen}, {Richards}, {Anderson}, \& {Schneider}}]{reye08}
{Reyes}, R., {Zakamska}, N.~L., {Strauss}, M.~A., {et~al.} 2008, \aj, 136, 2373

\bibitem[{{Rich} {et~al.}(2011){Rich}, {Kewley}, \& {Dopita}}]{rich11}
{Rich}, J.~A., {Kewley}, L.~J., \& {Dopita}, M.~A. 2011, \apj, 734, 87

\bibitem[{{Rupke} \& {Veilleux}(2011)}]{rupk11}
{Rupke}, D.~S.~N., \& {Veilleux}, S. 2011, \apjl, 729, L27

\bibitem[{{Sanders} {et~al.}(1988){Sanders}, {Soifer}, {Elias}, {Madore},
  {Matthews}, {Neugebauer}, \& {Scoville}}]{sand88}
{Sanders}, D.~B., {Soifer}, B.~T., {Elias}, J.~H., {et~al.} 1988, \apj, 325, 74

\bibitem[{{Scannapieco} \& {Oh}(2004)}]{scan04}
{Scannapieco}, E., \& {Oh}, S.~P. 2004, \apj, 608, 62

\bibitem[{{Schmitt} {et~al.}(2003){Schmitt}, {Donley}, {Antonucci},
  {Hutchings}, {Kinney}, \& {Pringle}}]{schm03}
{Schmitt}, H.~R., {Donley}, J.~L., {Antonucci}, R.~R.~J., {et~al.} 2003, \apj,
  597, 768

\bibitem[{{Shen} {et~al.}(2008){Shen}, {Greene}, {Strauss}, {Richards}, \&
  {Schneider}}]{shen08}
{Shen}, Y., {Greene}, J.~E., {Strauss}, M.~A., {Richards}, G.~T., \&
  {Schneider}, D.~P. 2008, \apj, 680, 169

\bibitem[{{Silk} \& {Rees}(1998)}]{silk98}
{Silk}, J., \& {Rees}, M.~J. 1998, \aap, 331, L1

\bibitem[{{Spinelli} {et~al.}(2006){Spinelli}, {Storchi-Bergmann}, {Brandt}, \&
  {Calzetti}}]{spin06}
{Spinelli}, P.~F., {Storchi-Bergmann}, T., {Brandt}, C.~H., \& {Calzetti}, D.
  2006, \apjs, 166, 498

\bibitem[{{Springel} {et~al.}(2005){Springel}, {Di Matteo}, \&
  {Hernquist}}]{spri05}
{Springel}, V., {Di Matteo}, T., \& {Hernquist}, L. 2005, \mnras, 361, 776

\bibitem[{{Steidel} {et~al.}(2010){Steidel}, {Erb}, {Shapley}, {Pettini},
  {Reddy}, {Bogosavljevi{\'c}}, {Rudie}, \& {Rakic}}]{stei10}
{Steidel}, C.~C., {Erb}, D.~K., {Shapley}, A.~E., {et~al.} 2010, \apj, 717, 289

\bibitem[{{Stockton} {et~al.}(2006){Stockton}, {Fu}, {Henry}, \&
  {Canalizo}}]{stoc06}
{Stockton}, A., {Fu}, H., {Henry}, J.~P., \& {Canalizo}, G. 2006, \apj, 638,
  635

\bibitem[{{Stockton} \& {MacKenty}(1987)}]{stoc87}
{Stockton}, A., \& {MacKenty}, J.~W. 1987, \apj, 316, 584

\bibitem[{{Tabor} \& {Binney}(1993)}]{tabo93}
{Tabor}, G., \& {Binney}, J. 1993, \mnras, 263, 323

\bibitem[{{Tadhunter}(1991)}]{tadh91}
{Tadhunter}, C.~N. 1991, \mnras, 251, 46P

\bibitem[{{Tchekhovskoy} {et~al.}(2010){Tchekhovskoy}, {Narayan}, \&
  {McKinney}}]{tche10}
{Tchekhovskoy}, A., {Narayan}, R., \& {McKinney}, J.~C. 2010, \apj, 711, 50

\bibitem[{{Thoul} \& {Weinberg}(1995)}]{thou95}
{Thoul}, A.~A., \& {Weinberg}, D.~H. 1995, \apj, 442, 480

\bibitem[{{Tran} {et~al.}(1995){Tran}, {Cohen}, \& {Goodrich}}]{tran95}
{Tran}, H.~D., {Cohen}, M.~H., \& {Goodrich}, R.~W. 1995, \aj, 110, 2597

\bibitem[{{Tremaine} {et~al.}(2002)}]{trem02}
{Tremaine}, S., {et~al.} 2002, \apj, 574, 740

\bibitem[{{Tremblay} {et~al.}(2009){Tremblay}, {Chiaberge}, {Sparks}, {Baum},
  {Allen}, {Axon}, {Capetti}, {Floyd}, {Macchetto}, {Miley}, {Noel-Storr},
  {O'Dea}, {Perlman}, \& {Quillen}}]{trem09}
{Tremblay}, G.~R., {Chiaberge}, M., {Sparks}, W.~B., {et~al.} 2009, \apjs, 183,
  278

\bibitem[{{van Breugel} {et~al.}(1986){van Breugel}, {Heckman}, {Miley}, \&
  {Filippenko}}]{vanb86}
{van Breugel}, W.~J.~M., {Heckman}, T.~M., {Miley}, G.~K., \& {Filippenko},
  A.~V. 1986, \apj, 311, 58

\bibitem[{{van Dokkum}(2001)}]{vdok01}
{van Dokkum}, P.~G. 2001, \pasp, 113, 1420

\bibitem[{{Veilleux} {et~al.}(2005){Veilleux}, {Cecil}, \&
  {Bland-Hawthorn}}]{veil05}
{Veilleux}, S., {Cecil}, G., \& {Bland-Hawthorn}, J. 2005, \araa, 43, 769

\bibitem[{{Veilleux} \& {Osterbrock}(1987)}]{veil87}
{Veilleux}, S., \& {Osterbrock}, D.~E. 1987, \apjs, 63, 295

\bibitem[{{Vignali} {et~al.}(2010){Vignali}, {Alexander}, {Gilli}, \&
  {Pozzi}}]{vign10}
{Vignali}, C., {Alexander}, D.~M., {Gilli}, R., \& {Pozzi}, F. 2010, \mnras,
  404, 48

\bibitem[{{Villar-Mart{\'{\i}}n}
  {et~al.}(2011{\natexlab{a}}){Villar-Mart{\'{\i}}n}, {Humphrey}, {Delgado},
  {Colina}, \& {Arribas}}]{vill11b}
{Villar-Mart{\'{\i}}n}, M., {Humphrey}, A., {Delgado}, R.~G., {Colina}, L., \&
  {Arribas}, S. 2011{\natexlab{a}}, \mnras, 418, 2032

\bibitem[{{Villar-Mart{\'{\i}}n} {et~al.}(2008){Villar-Mart{\'{\i}}n},
  {Humphrey}, {Mart{\'{\i}}nez-Sansigre}, {P{\'e}rez-Torres}, {Binette}, \&
  {Zhang}}]{vill08}
{Villar-Mart{\'{\i}}n}, M., {Humphrey}, A., {Mart{\'{\i}}nez-Sansigre}, A.,
  {et~al.} 2008, \mnras, 390, 218

\bibitem[{{Villar-Mart{\'{\i}}n}
  {et~al.}(2011{\natexlab{b}}){Villar-Mart{\'{\i}}n}, {Tadhunter}, {Humphrey},
  {Encina}, {Delgado}, {Torres}, \& {Mart{\'{\i}}nez-Sansigre}}]{vill11a}
{Villar-Mart{\'{\i}}n}, M., {Tadhunter}, C., {Humphrey}, A., {et~al.}
  2011{\natexlab{b}}, \mnras, 416, 262

\bibitem[{{Villar-Mart{\'{\i}}n} {et~al.}(1999){Villar-Mart{\'{\i}}n},
  {Tadhunter}, {Morganti}, {Axon}, \& {Koekemoer}}]{vill99}
{Villar-Mart{\'{\i}}n}, M., {Tadhunter}, C., {Morganti}, R., {Axon}, D., \&
  {Koekemoer}, A. 1999, \mnras, 307, 24

\bibitem[{{Voit}(2005)}]{voit05a}
{Voit}, G.~M. 2005, Reviews of Modern Physics, 77, 207

\bibitem[{{Voit} \& {Donahue}(2005)}]{voit05b}
{Voit}, G.~M., \& {Donahue}, M. 2005, \apj, 634, 955

\bibitem[{{Weymann} {et~al.}(1981){Weymann}, {Carswell}, \& {Smith}}]{weym81}
{Weymann}, R.~J., {Carswell}, R.~F., \& {Smith}, M.~G. 1981, \araa, 19, 41

\bibitem[{{White} {et~al.}(1997){White}, {Becker}, {Helfand}, \&
  {Gregg}}]{whit97}
{White}, R.~L., {Becker}, R.~H., {Helfand}, D.~J., \& {Gregg}, M.~D. 1997,
  \apj, 475, 479

\bibitem[{{Whittle}(1992)}]{whit92}
{Whittle}, M. 1992, \apjs, 79, 49

\bibitem[{{Xu} {et~al.}(1999){Xu}, {Livio}, \& {Baum}}]{xu99}
{Xu}, C., {Livio}, M., \& {Baum}, S. 1999, \aj, 118, 1169

\bibitem[{{York} {et~al.}(2000)}]{york00}
{York}, D.~G., {et~al.} 2000, \aj, 120, 1579

\bibitem[{{Zakamska} {et~al.}(2008){Zakamska}, {G{\'o}mez}, {Strauss}, \&
  {Krolik}}]{zaka08}
{Zakamska}, N.~L., {G{\'o}mez}, L., {Strauss}, M.~A., \& {Krolik}, J.~H. 2008,
  \aj, 136, 1607

\bibitem[{{Zakamska} {et~al.}(2004){Zakamska}, {Strauss}, {Heckman},
  {Ivezi{\'c}}, \& {Krolik}}]{zaka04}
{Zakamska}, N.~L., {Strauss}, M.~A., {Heckman}, T.~M., {Ivezi{\'c}}, {\v Z}.,
  \& {Krolik}, J.~H. 2004, \aj, 128, 1002

\bibitem[{{Zakamska} {et~al.}(2003)}]{zaka03}
{Zakamska}, N.~L., {et~al.} 2003, \aj, 126, 2125

\bibitem[{{Zakamska} {et~al.}(2005)}]{zaka05}
---. 2005, \aj, 129, 1212

\bibitem[{{Zakamska} {et~al.}(2006)}]{zaka06}
---. 2006, \aj, 132, 1496

\bibitem[{{Zubovas} \& {King}(2012)}]{zubo12}
{Zubovas}, K., \& {King}, A. 2012, \apjl, 745, L34

\end{thebibliography}




\clearpage


\end{document}